\documentclass[a4paper,fleqn,usenatbib]{mnras}
\usepackage{newtxtext,newtxmath}
\usepackage[T1]{fontenc}
\usepackage{ae,aecompl}
\usepackage{deluxetable}
\usepackage{color}
\usepackage{graphicx}	% Including figure files
\usepackage{amsmath}	% Advanced maths commands
\usepackage{amssymb}	% Extra maths symbols
\usepackage{natbib}
\title[Gaia16apd]{Gaia16apd -- a link between fast-and slowly-declining type I superluminous supernovae}

\author[T. Kangas et al.]{T. Kangas$^{1}$\thanks{E-mail: tjakan@utu.fi}, 
N. Blagorodnova$^{2}$, 
S. Mattila$^{1}$, 
P. Lundqvist$^{3}$, 
M. Fraser$^{4,5}$,  \and
U. Burgaz$^{6}$,
E. Cappellaro$^{7}$,
J.~M. Carrasco Mart\'{i}nez$^{8}$,
N. Elias-Rosa$^{7}$,
L. K. Hardy$^{9}$, \and
J. Harmanen$^{1}$, 
E. Y. Hsiao$^{10}$, 
J. Isern$^{11,12}$,
E. Kankare$^{13}$,
Z. Ko{\l}aczkowski$^{14}$, \and
M. B. Nielsen$^{15}$,
T. M. Reynolds$^{1,16}$, 
L. Rhodes$^{9}$, 
A. Somero$^{1}$ 
M. D. Stritzinger$^{15}$, \and
and \L{}. Wyrzykowski$^{17}$
\\
$^{1}$Tuorla Observatory, Department of Physics and Astronomy, University of Turku, V\"{a}is\"{a}l\"{a}ntie 20, FI-21500 Piikki\"{o}, Finland\\
$^{2}$Cahill Center for Astrophysics, California Institute of Technology, Pasadena, CA 91125, USA\\
$^{3}$Department of Astronomy and The Oskar Klein Centre, AlbaNova University Center, Stockholm University, SE-10691 Sweden\\
$^{4}$Institute of Astronomy, University of Cambridge, Madingley Road, Cambridge CB3 0HA, United Kingdom\\
$^{5}$School of Physics, O'Brien Centre for Science North, University College Dublin, Belfield, Dublin 4, Ireland\\
$^{6}$Department of Astronomy and Space Sciences, Ege University, 35100, Izmir, Turkey\\
$^{7}$INAF - Osservatorio Astronomico di Padova, Vicolo dell'Osservatorio 5, Padova I-35122, Italy\\
$^{8}$Institut de Ci\'{e}ncies del Cosmos, Universitat de Barcelona (IEEC-UB), Mart\'{i} Franqu\'{e}s 1, E-08028 Barcelona, Spain\\
$^{9}$Department of Physics and Astronomy, University of Sheffield, Sheffield, S3 7RH, UK\\
$^{10}$Department of Physics, Florida State University, 77 Chieftain Way, Tallahassee, FL 32306, USA\\
$^{11}$Institut de Ci\'{e}ncies de l'Espai (CSIC), Campus UAB, 08193 Cerdanyola, Spain\\
$^{12}$Institut d'Estudis Espacials de Catalunya, Ed. Nexus-201, c/Gran Capit\'{a} 2-4, 08034 Barcelona, Spain\\
$^{13}$Astrophysics Research Centre, School of Mathematics and Physics, Queen's University Belfast, Belfast BT7 1NN, UK\\
$^{14}$Instytut Astronomiczny, Uniwersytet Wroc{\l}awski, Kopernika 11, 51-622, Wroc{\l}aw, Poland\\
$^{15}$Department of Physics and Astronomy, Aarhus University, Ny Munkegade 120, DK-8000 Aarhus C, Denmark \\
$^{16}$Nordic  Optical  Telescope,  Apartado  474,  E-38700  Santa Cruz de La Palma, Spain \\
$^{17}$Warsaw University Astronomical Observatory, Al. Ujazdowskie 4, 00-478 Warszawa, Poland
}

\date{Accepted XXX. Received YYY; in original form ZZZ}

\pubyear{2017}

\begin{document}
\label{firstpage}
\pagerange{\pageref{firstpage}--\pageref{lastpage}}
\maketitle

\begin{abstract}
We present ultraviolet, optical and infrared photometry and optical spectroscopy of the type Ic superluminous supernova (SLSN) Gaia16apd (= SN~2016eay), covering its evolution from 26 d before the $g$-band peak to 234.1 d after the peak. Gaia16apd was followed as a part of the NOT Unbiased Transient Survey (NUTS). It is one of the closest SLSNe known ($z = 0.102\pm0.001$), with detailed optical and ultraviolet (UV) observations covering the peak. Gaia16apd is a spectroscopically typical type Ic SLSN, exhibiting the characteristic blue early spectra with O {\sc ii} absorption, and reaches a peak $M_{g} = -21.8 \pm 0.1$ mag. However, photometrically it exhibits an evolution intermediate between the fast- and slowly-declining type Ic SLSNe, with an early evolution closer to the fast-declining events. Together with LSQ12dlf, another SLSN with similar properties, it demonstrates a possible continuum between fast- and slowly-declining events. It is unusually UV-bright even for a SLSN, reaching a non-$K$-corrected $M_{uvm2} \simeq -23.3$ mag, the only other type Ic SLSN with similar UV brightness being SN~2010gx. Assuming that Gaia16apd was powered by magnetar spin-down, we derive a period of $P = 1.9\pm0.2$ ms and a magnetic field of $B = 1.9\pm0.2 \times 10^{14}$ G for the magnetar. The estimated ejecta mass is between 8 and 16 $\mathrm{M}_{\odot}$ and the kinetic energy between 1.3 and $2.5 \times 10^{52}$ erg, depending on opacity and assuming that the entire ejecta is swept up into a thin shell. Despite the early photometric differences, the spectra at late times are similar to slowly-declining type Ic SLSNe, implying that the two subclasses originate from similar progenitors.
\end{abstract}

%The bolometric peak luminosity is $2.8 \pm 0.1 \times 10^{44}$ erg s$^{-1}$.


%%%%%%%%%%%%%%%%%%%%%%%%%%%%%%%%%%%%%%%%%%%%%%%%%%

%%%%%%%%%%%%%%%%% BODY OF PAPER %%%%%%%%%%%%%%%%%%

\section{Introduction} \label{sec:intro}

Superluminous supernovae (SLSNe) are explosions of massive stars that reach peak absolute magnitudes $\le -21$ mag \citep[e.g.][]{quimby11,galyam12}, making them up to hundreds of times brighter than normal supernovae (SNe). Like normal core-collapse supernovae (CCSNe), SLSNe are divided into hydrogen-rich (type II) and hydrogen-poor (type I) events based on their spectroscopy. Some SLSNe evolve to spectroscopically resemble normal type Ic SNe a few weeks after maximum light \citep[e.g.][]{pastorello10,inserra13} and are called type Ic SLSNe. They are rare events, with an estimated rate of 0.01 per cent of the CCSN rate \citep{quimby13,mccrum15}. The host galaxies of type Ic SLSNe are typically faint and metal-poor \citep[e.g.][]{chen13,chen15,perley16}, and low metallicity has been suggested to be necessary for their progenitors. Type Ic SLSNe include both slowly declining events such as SN~2007bi \citep{galyam10} or PTF12dam \citep{nicholl13}, whose decline rates initially resemble $^{56}$Co decay \citep[e.g.][]{inserra17}, and significantly faster-declining events such as SN~2005ap \citep{quimby07}, SN~2010gx \citep{pastorello10} or SN~2011ke \citep{inserra13}. The $e$-folding decline time-scales of fast- and slowly-declining events appear to be clustered around 30 d and 70 d respectively, but this bimodality may not be significant \citep{nicholl15}.

Different power sources have been suggested to account for the high luminosity of SLSNe. Slowly declining SN~2007bi-like events \citep[e.g.][]{galyam10} have been suggested to be pair-instability SNe \citep[PISNe;][]{hegerwoosley02}, where an extremely massive star is completely disrupted in a thermonuclear runaway process. However, in recent studies \citep[e.g.][]{dessart13,mccrum14,lunnan16} such events have been found incompatible with PISN models \citep{kasen11}. The decay of $^{56}$Ni cannot be the primary power source, as it fails to produce both the required peak luminosity and the tail-phase light curve self-consistently \citep{quimby11}. Interaction with a circumstellar medium (CSM) remains a plausible way to power at least some SLSNe \citep[e.g.][]{chevalierirwin11}; however, fine-tuning of the CSM mass and density is required in order to explain some of the observational properties of type Ic SLSNe \citep{nicholl15}. Furthermore, type Ic SLSNe do not show signs of CSM interaction in their spectra \citep[e.g.][]{nicholl14}. The currently dominant explanation for the luminosity is a central engine such as the spin-down of a millisecond magnetar with $B \sim 10^{14}$ G \citep{kasenbildsten10,woosley10} or possibly accretion onto a black hole from fall-back of ejected material \citep{dexterkasen13}. \citet{inserra17} did, however, find that interaction with a small amount of CSM may be present in slowly-declining type Ic SLSNe, even though they favoured the magnetar engine as the primary power source.

In this paper, we present photometric and spectroscopic observations and analysis of the type Ic SLSN Gaia16apd. We show that, early on, this event resembles the archetypal fast-declining type Ic SLSN, SN~2010gx. Later it evolves in a way intermediate between the fast- and slowly-declining type Ic SLSNe. Gaia16apd exhibits an ultraviolet (UV) brightness much higher than slowly-declining type Ic SLSNe, and is one of a scant few other type Ic SLSNe so far with good UV sampling, highlighting the necessity of UV observations for understanding the variation inside this class of SLSNe. We use magnetar light curve models to estimate physical parameters for the explosion and compare the late-time (151.6 to 234.1 d after maximum light) spectra to the slowly-declining events, showing a striking similarity despite the differences in the early photometric evolution.

\section{Observations and data reduction} \label{sec:data}

Throughout the paper, we assume a flat $\Lambda$CDM cosmology with $H_0 = 69.6$ km s$^{-1}$ Mpc$^{-1}$, $\Omega_M = 0.286$ and $\Omega_{\Lambda} = 0.714$ \citep{cosmology}. The redshift of the SN ($z = 0.102 \pm 0.001$; see Section \ref{sec:host}) corresponds to a luminosity distance $d_L = 473.3^{+5.0}_{-4.9}$ Mpc and a distance modulus of $\mu = 38.38 \pm 0.03$. A Galactic reddening of $E(B-V)_{\mathrm{gal}} = 0.0132$ mag was adopted using the Galactic dust maps by \citet{dustmap}.

\subsection{Discovery and classification}

Gaia16apd (= SN~2016eay) was discovered by the \emph{Gaia} Photometric Science Alerts programme\footnote{http://gsaweb.ast.cam.ac.uk/alerts/} of the \emph{Gaia} satellite \citep{gaia} operated by the European Space Agency (ESA) on 2016 May 16.80 UT (MJD = 57524.80) at $\alpha = 12^{\mathrm{h}} 02^{\mathrm{m}} 51\fs71, \delta = +44\degr 15\arcmin 27\farcs40$ (J2000.0) with a brightness of 17.35 mag in the \emph{Gaia} $G$-band. The discovery was reported on the \emph{Gaia} Photometric Science Alerts website on May 18. The last \emph{Gaia} non-detection, with a limiting magnitude of $\geq 20.5$ mag, is from April 5.2 UT. Later $g$-band non-detections on April 9.7 $\pm$ 8.5 d (co-added) and April 18.3, down to limiting magnitudes of 22.1 mag and 21.0 mag respectively, were made by the Intermediate Palomar Transient Factory\footnote{http://www.ptf.caltech.edu/iptf} (iPTF). The target was recovered in images taken with the Palomar 60-inch Telescope (P60) on May 12.3 (MJD 57520.32) at $g = 17.3 \pm 0.2$ mag \citep{yan16}. \citet{classif} determined the redshift of Gaia16apd to be $z = 0.102$ using host galaxy lines -- resulting in an absolute discovery magnitude of $\sim-21.0$ mag -- and classified the event as a type I SLSN using the 2.56-m Nordic Optical Telescope \citep[NOT;][]{notref} as part of the NOT Un-biased Transient Survey (NUTS)\footnote{csp2.lco.cl/not} collaboration \citep[][]{nuts}.

\subsection{Host galaxy} \label{sec:host}

\begin{figure*}
\begin{minipage}{1.\linewidth}
\centering
\includegraphics[width=1\columnwidth]{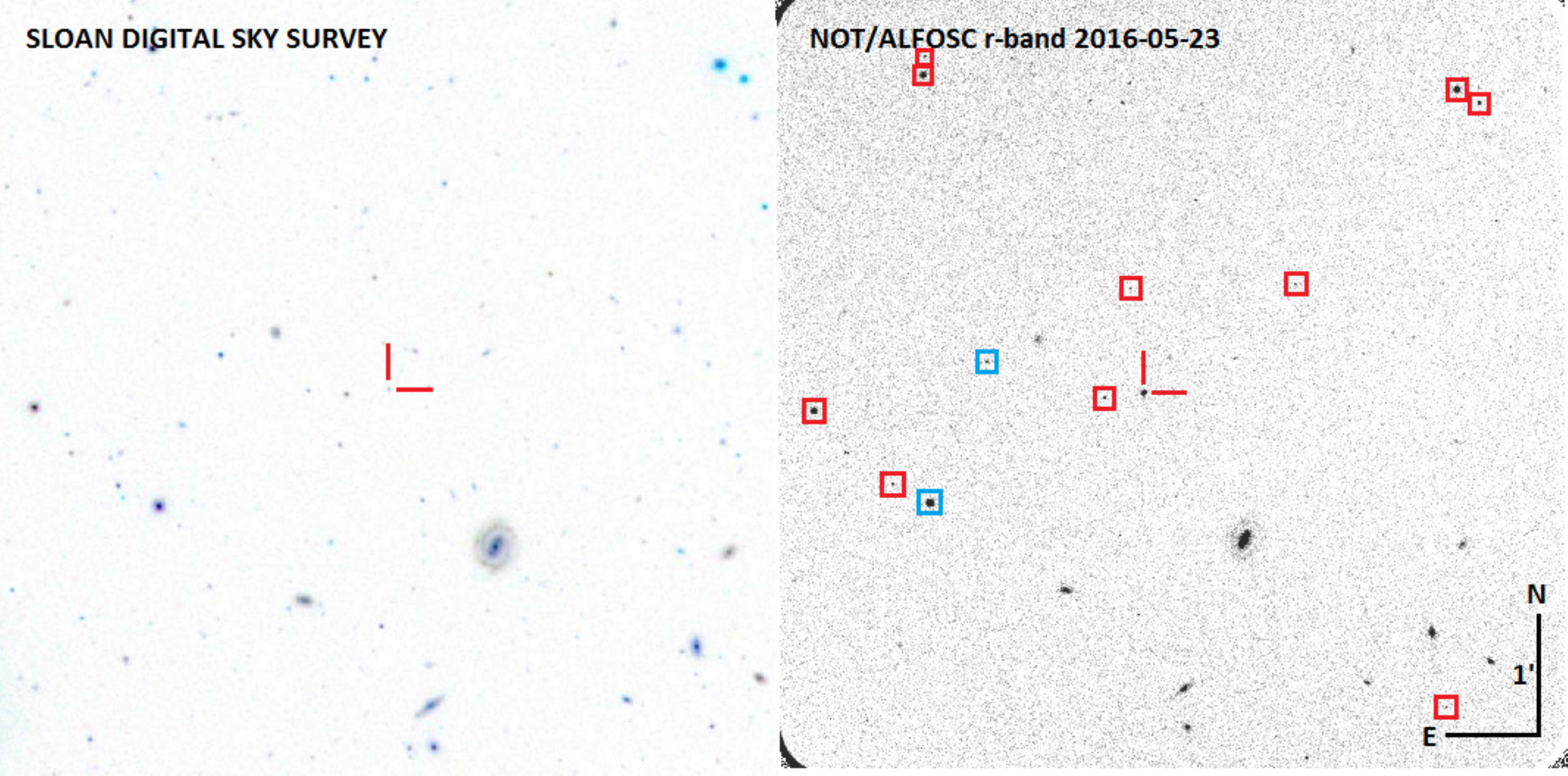}
\caption{SDSS image of the field of Gaia16apd (left) and our first $r$-band image from NOT/ALFOSC (right). The position of Gaia16apd is indicated with red tick marks in both images, and the stars used in the photometric calibration of the NOT images have been marked with squares. The two 2MASS stars in the field are marked with blue squares. These two stars were used to calibrate both the optical and the NIR images. The stars marked with red squares were only used to calibrate the optical images.}
\label{fig:fc}
\end{minipage}
\end{figure*}

\begin{figure*}
\begin{minipage}{1.\linewidth}
\centering
\includegraphics[width=1\columnwidth]{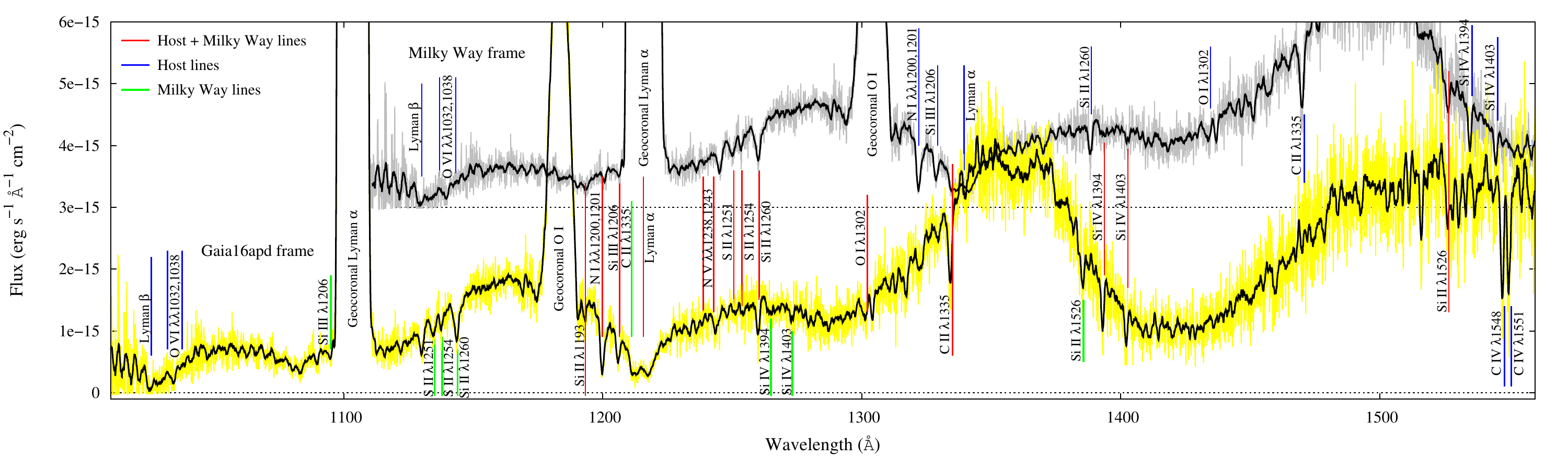}
\caption{\emph{HST}/COS spectrum of Gaia16apd \citep{yan16} at the rest frame of the SN (yellow) and in the Milky Way rest frame (i.e. the observed spectrum; grey). The black spectra were obtained by Savitzky-Golay smoothing. The observed spectrum was shifted to facilitate comparisons. Spectral features indicated by red lines are formed both in the host galaxy of the SN and in the Milky Way. Green lines indicate features originating in the Milky Way, and blue is for features originating in the host galaxy. In general, the line identifications are similar to those in \citet{yan16}.}
\label{fig:cos}
\end{minipage}
\end{figure*}

\begin{figure}
\centering
\includegraphics[width=1\columnwidth]{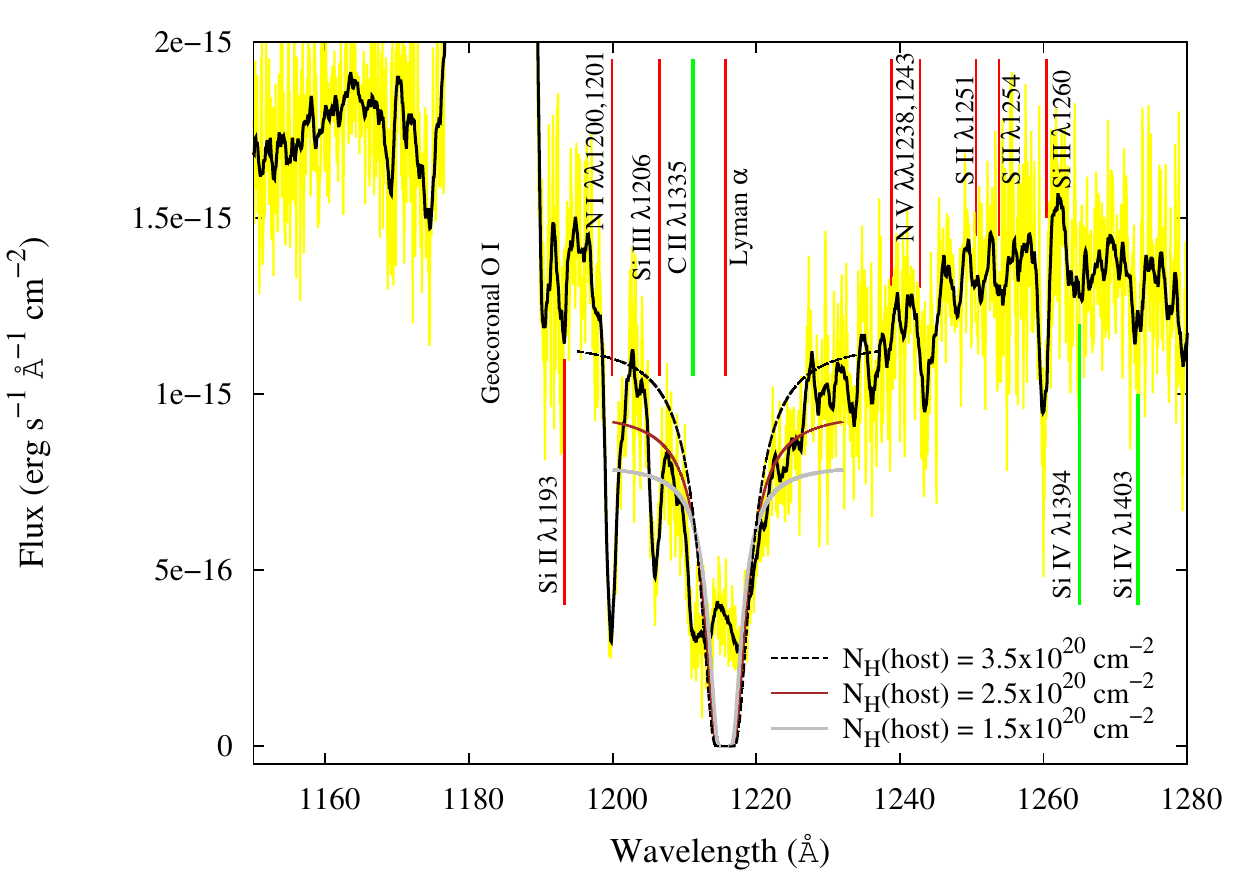}
\caption{As Figure \ref{fig:cos}, but zoomed in to the Ly~$\alpha$ region used to estimate the column density of neutral hydrogen in the host galaxy. We have overlaid modeled absorption profiles for L~$\alpha$ for three column densities of neutral hydrogen in the host, $N_{\rm H}$: $1.5\times10^{20}$~cm$^{-2}$ (grey), $2.5\times10^{20}$~cm$^{-2}$ (brown) and $3.5\times10^{20}$~cm$^{-2}$ (black dashed).}
\label{fig:cos2}
\end{figure}

Figure \ref{fig:fc} shows the location of the SN and its faint host galaxy in our first $r$-band image from the NOT using the Andalucia Faint Object Spectrograph and Camera (ALFOSC) instrument and a pre-explosion Sloan Digital Sky Survey (SDSS) image, with the field stars that have been used for calibrating the NOT images (see Section \ref{sec:phot}). According to SDSS Data Release 13 \citep{sdss13}, the $g$-band magnitude of the host galaxy SDSS J120251.71+441527.4 is $21.73 \pm 0.06$ mag. Narrow, persistent Balmer emission lines and [O {\sc iii}] $\lambda\lambda4959,5007$ attributed to the host galaxy are visible in each spectrum of Gaia16apd (see Section \ref{sec:spec}). Using these narrow lines, we derived a redshift of $z = 0.102 \pm 0.001$, which was adopted for Gaia16apd. This redshift is consistent with the values of 0.1018 and 0.1013 reported by \citet{yan16} and \citet{nicholl16b}, respectively. Thus the non-$K$-corrected absolute magnitude is $M_{g} = -16.69 \pm 0.07$ mag, consistent with the faint dwarf galaxies that typically host type I SLSNe \citep{leloudas15,lunnan15,perley16}. We estimated an upper limit for the [N {\sc ii}] $\lambda6583$ / H\,$\alpha$ ratio as log [N {\sc ii}]$\lambda6583$/H\,$\alpha$ $\lesssim -0.80$ dex. The measured average [O {\sc iii}] $\lambda$5007 / H\,$\beta$ ratio is log [O {\sc iii}]$\lambda$5007/H\,$\beta$ = $0.68 \pm 0.05$ dex. These values are again consistent with other type I SLSN host galaxies. Thus the host galaxy of Gaia16apd seems typical for a type Ic SLSN. Using the $R_{23}$ diagnostic \citep{kobulnicky99}, we obtain an estimated metallicity of $0.2~Z_{\odot}$, in agreement with \citet{nicholl16b}.

No Na {\sc i} D absorption lines from the ISM in the host galaxy were detected in the optical spectra. However, a \emph{Hubble Space Telescope} (\emph{HST}) spectrum from MJD 57541.2 (7.9 d before $g$-band peak) observed with the COS instrument, published by \citet{yan16}, was used to roughly estimate the host galaxy extinction. The calibrated spectrum (Figure \ref{fig:cos}) was obtained from the \emph{HST} archive. Several spectral lines formed in the host galaxy of the SN and in the Milky Way are noted. In general, the host lines are somewhat stronger than lines formed in the Milky Way -- particularly for the high-ionisation doublets C~{\sc iv} $\lambda\lambda$1548,1551 and Si~{\sc iv} $\lambda\lambda$1394,1403, and to some extent Si~{\sc iii} $\lambda$1206. A noticeable difference in strength is also noted for N~{\sc i}~$\lambda\lambda$1200,1201, where the absorption in the host is much stronger than in the Milky Way, possibly indicating enhanced nitrogen abundance in the host. S~{\sc ii} and Si~{\sc ii} lines are more similar in strength for the two galaxies. In Figure \ref{fig:cos2}, we have overlaid modeled absorption profiles for L~$\alpha$ for three column densities of neutral hydrogen in the host, $N_{\rm H}$, namely $1.5\times10^{20}$~cm$^{-2}$, $2.5\times10^{20}$~cm$^{-2}$ and $3.5\times10^{20}$~cm$^{-2}$. Although the baseline for the absorption line is not well defined, the fit for L$\alpha$ suggests a column density within this range -- we adopt $2.5 \pm 1.0 \times10^{20}$~cm$^{-2}$ as our estimate. Using the relation between $E(B-V)$ and $N_{\rm H}$ of \citet{guver09}, this would result in $E(B-V)_{\mathrm{host}} = 0.04 \pm 0.02$ mag assuming Solar metallicity. At the metallicity of the host, $0.2~Z_{\odot}$, $E(B-V)_{\mathrm{host}}\sim0.01$ mag would be closer to the truth, considering the typical gas-to-dust mass ratio as a function of metallicity \citep{remy14}. Thus we adopt an estimated $E(B-V)_{\mathrm{host}} = 0.010 \pm 0.005$ mag, a value comparable to the foreground $E(B-V)_{\mathrm{gal}} = 0.0132$ mag. This results in a non-negligible correction in the UV, e.g. $0.09 \pm 0.05$ mag in $uvm2$.

\subsection{Photometry} \label{sec:phot}

Optical and near-infrared (NIR) photometric follow-up observations were performed using the NOT as a part of the NUTS programme. Optical imaging in the \emph{UBVri} bands was obtained using ALFOSC, while NIR imaging in the \emph{JHKs} bands was obtained with the NOT near-infrared Camera and spectrograph (NOTCam) instrument. Optical photometric observations in the \emph{RI} bands were also performed using the 0.5-m robotic telescope \emph{pt5m} \citep{pt5m} on La Palma, Canary Islands, and \emph{VRI}-band observations using the 0.8-m robotic Joan Or\'{o} Telescope (TJO) at the Montsec Astronomical Observatory (OAdM\footnote{http://www.oadm.cat}), Lleida, Spain. UV and $UBV$-band photometry was performed with the Ultraviolet/Optical Telescope (UVOT) aboard \emph{Swift} \citep[the first \emph{Swift} observations are reported in a telegram by][]{swift_atel}. We also included the public \emph{Gaia} photometry points, converted to $V$-band according to \citet{jordi10}, and the iPTF pre-discovery $g$-band point \citep{yan16}. Gaia16apd was observable until early August 2016, at which point it was obscured by the Sun. Observations resumed at the end of October 2016.

Reduction of the optical NOT images was done using the custom pipeline {\sc foscgui}\footnote{http://sngroup.oapd.inaf.it/foscgui.html}. The NIR data were reduced using a slightly modified version of the NOTCam Quicklook v2.5 reduction package\footnote{http://www.not.iac.es/instruments/notcam/guide/observe. html\#reductions}. Both use standard {\sc iraf}\footnote{{\sc iraf} is distributed by the National Optical Astronomy Observatory, which is operated by the Association of Universities for Research in Astronomy (AURA) under cooperative agreement with the National Science Foundation.} tasks. The zero points of the \emph{ri} images were calibrated relative to 12 nearby stars with magnitudes available at the SDSS Data Release 13 SkyServer\footnote{http://skyserver.sdss.org/dr13/en/home.aspx}. The \emph{UBV} magnitudes of the same field stars were calibrated using the standard star fields GD 246 and PG 1047 observed on the photometric nights of 2016 July 13 and 2017 January 26, respectively. The zero points of the \emph{JHKs} images were calibrated relative to two nearby Two Micron All-Sky Survey \citep[2MASS;][]{skrutskie06}\footnote{http://www.ipac.caltech.edu/2mass/} stars. The photometry was performed using the {\sc SNOoPy} package\footnote{http://sngroup.oapd.inaf.it/snoopy.html} inside the {\sc quba} pipeline \citep{quba}, which fits the point spread function (PSF) using the {\sc iraf} package {\sc daophot}. The residuals from the PSF fits were minimal, indicating that a single point source is a good approximation for the SN and the host galaxy combined. Images from \emph{pt5m} were reduced by an automatic pipeline that performs bias and flat-field correction and uses the {\sc SExtractor} package \citep{bertin96} to extract magnitudes. Images from TJO were bias-, dark- and flat-field-corrected using the {\sc ccdproc} package in {\sc astropy} \citep{craig15}. Astrometry was performed using Astrometry.net \citep{lang10}, and instrumental magnitudes obtained using {\sc SExtractor}. Preliminary photometry from \emph{pt5m} and TJO was collected at the Cambridge Photometric Calibration Server\footnote{http://gsaweb.ast.cam.ac.uk/followup/}. Final PSF photometry and colour-term calibrations were performed with standard {\sc iraf} tasks using local SDSS stars and the conversion formulae by \citet{jordi06} to obtain \emph{gri} magnitudes. \emph{Swift} photometry was reduced using the {\sc uvotsource} task in the {\sc HEAsoft} package. 

\begin{table*}
\begin{minipage}{0.92\linewidth}
\begin{small}
\caption{Log of optical photometric observations of Gaia16apd. SDSS magnitudes are reported in the AB system, while other magnitudes are in the Vega system. No $K$-corrections have been applied. The host galaxy magnitudes (from SDSS) have been subtracted from points marked with an asterisk (*).}
\centering
\begin{tabular}{lcccccccc}
\hline
Phase\footnote{Rest-frame days since absolute $g$-band peak at MJD 57549.1.} & MJD & Telescope & $U$ & $B$ & $g$ & $V$ & $r$ & $i$ \\
(d) &  &  & (mag) & (mag) & (mag) & (mag) & (mag) & (mag) \\
\hline
$-22.1$ & 57524.8 & \emph{Gaia} & -- & -- & -- & 17.39$\pm$0.06 & -- & -- \\
$-22.0$ & 57524.9 & \emph{Gaia} & -- & -- & -- & 17.35$\pm$0.07 & -- & -- \\
$-18.3$ & 57528.9 & TJO & -- & -- & -- & -- & 17.31$\pm$0.01 & 17.50$\pm$0.02 \\
$-17.6$ & 57529.7 & \textit{Swift} & 15.62$\pm$0.03 & 16.95$\pm$0.04 & -- & 16.95$\pm$0.08 & -- & -- \\
$-15.8$ & 57531.7 & \textit{Swift} & 15.61$\pm$0.04 & 16.91$\pm$0.06 & -- & 16.90$\pm$0.10 & -- & -- \\
$-15.5$ & 57532.0 & NOT & 15.56$\pm$0.19 & 16.76$\pm$0.03 & -- & 16.91$\pm$0.03 & 17.20$\pm$0.08 & 17.31$\pm$0.08 \\
$-14.7$ & 57532.9 & \emph{pt5m} & -- & -- & -- & -- & 17.08$\pm$0.03 & -- \\
$-14.5$ & 57533.2 & \textit{Swift} & 15.43$\pm$0.05 & 16.70$\pm$0.07 & -- & 16.75$\pm$0.12 & -- & -- \\
$-13.7$ & 57534.0 & \emph{pt5m} & -- & -- & -- & -- & 16.99$\pm$0.03 & 17.21$\pm$0.08 \\
$-12.9$ & 57534.9 & \emph{pt5m} & -- & -- & -- & -- & 16.98$\pm$0.02 & -- \\
$-12.0$ & 57535.9 & NOT & -- & 16.77$\pm$0.03 & -- & 16.79$\pm$0.03 & 17.01$\pm$0.04 & 17.13$\pm$0.04 \\
$-12.0$ & 57535.9 & TJO & -- & -- & 16.59$\pm$0.01 & -- & 16.98$\pm$0.01 & 17.15$\pm$0.01 \\
$-11.1$ & 57536.9 & \emph{pt5m} & -- & -- & -- & -- & 16.89$\pm$.02 & -- \\
$-10.6$ & 57537.5 & \textit{Swift} & 15.38$\pm$0.04 & 16.75$\pm$0.05 & -- & 16.61$\pm$0.08 & -- & -- \\
$-9.3$ & 57538.9 & \emph{pt5m} & -- & -- & -- & -- & 16.75$\pm$0.05 & -- \\
$-8.3$ & 57539.9 & \emph{pt5m} & -- & -- & -- & -- & 16.84$\pm$0.03 & -- \\
$-7.4$ & 57540.9 & \emph{pt5m} & -- & -- & -- & -- & 16.78$\pm$0.02 & 17.00$\pm$0.04 \\
$-6.9$ & 57541.5 & \textit{Swift} & 15.26$\pm$0.04 & 16.60$\pm$0.05 & -- & 16.52$\pm$0.08 & -- & -- \\
$-6.5$ & 57541.9 & TJO & -- & -- & 16.47$\pm$0.01 & -- & 16.78$\pm$0.01 & 16.95$\pm$0.01 \\
$-6.5$ & 57542.0 & NOT & 15.20$\pm$0.08 & 16.68$\pm$0.04 & -- & 16.61$\pm$0.04 & 16.81$\pm$0.04 & 16.90$\pm$0.04 \\
$-5.6$ & 57542.9 & \emph{pt5m} & -- & -- & -- & -- & 16.66$\pm$0.02 & -- \\
$-5.2$ & 57543.3 & \textit{Swift} & 15.31$\pm$0.04 & 16.52$\pm$0.05 & -- & 16.44$\pm$0.08 & -- & -- \\
$-4.7$ & 57543.9 & \emph{pt5m} & -- & -- & -- & -- & 16.70$\pm$0.02 & 16.83$\pm$0.06 \\
$-3.8$ & 57544.9 & \emph{pt5m} & -- & -- & -- & -- & 16.69$\pm$0.02 & 16.89$\pm$0.05 \\
$-3.5$ & 57545.3 & \textit{Swift} & 15.33$\pm$0.04 & 16.52$\pm$0.05 & -- & 16.57$\pm$0.08 & -- & -- \\
$-2.9$ & 57545.9 & \emph{pt5m} & -- & -- & -- & -- & 16.67$\pm$0.02 & -- \\
$-2.0$ & 57546.9 & \emph{pt5m} & -- & -- & -- & -- & 16.55$\pm$0.02 & -- \\
$-1.6$ & 57547.3 & \textit{Swift} & 15.31$\pm$0.03 & 16.51$\pm$0.04 & -- & 16.46$\pm$0.07 & -- & -- \\
$-1.0$ & 57548.0 & TJO & -- & -- & 16.41$\pm$0.01 & -- & 16.68$\pm$0.01 & 16.83$\pm$0.01 \\
1.3 & 57550.6 & \textit{Swift} & 15.34$\pm$0.04 & 16.50$\pm$0.04 & -- & 16.52$\pm$0.08 & -- & --\\
1.7 & 57550.9 & TJO & -- & -- & -- & -- & 16.71$\pm$0.01 & 16.84$\pm$0.01 \\
3.5 & 57553.0 & NOT & 15.74$\pm$0.05 & 16.61$\pm$0.14 & -- & 16.46$\pm$0.04 & 16.77$\pm$0.08 & 16.85$\pm$0.08 \\
3.8 & 57553.3 & \textit{Swift} & 15.52$\pm$0.06 & 16.57$\pm$0.07 & -- & -- & -- & -- \\
5.2 & 57554.9 & \emph{Gaia} & -- & -- & -- & 16.64$\pm$0.12 & -- & -- \\
6.2 & 57555.9 & \emph{pt5m} & -- & -- & -- & -- & 16.70$\pm$0.04 & -- \\
6.2 & 57555.9 & TJO & -- & -- & 16.55$\pm$0.01 & -- & 16.74$\pm$0.01 & 16.89$\pm$0.01 \\
7.1 & 57556.9 & \textit{Swift} & 15.62$\pm$0.04 & 16.59$\pm$0.05 & -- & 16.51$\pm$0.08 & -- & -- \\
9.3 & 57559.4 & \textit{Swift} & 15.67$\pm$0.04 & 16.55$\pm$0.05 & -- & 16.42$\pm$0.09 & -- & -- \\
11.8 & 57562.1 & \textit{Swift} & 15.78$\pm$0.05 & 16.73$\pm$0.05 & -- & 16.61$\pm$0.08 & -- & -- \\
13.5 & 57564.0 & NOT & 15.92$\pm$0.12 & 16.86$\pm$0.05 & -- & 16.72$\pm$0.05 & 16.84$\pm$0.05 & 16.94$\pm$0.05 \\
16.2 & 57566.9 & \emph{pt5m} & -- & -- & -- & -- & 16.85$\pm$0.02 & 16.95$\pm$0.08 \\
18.9 & 57569.9 & \emph{pt5m} & -- & -- & -- & -- & 16.94$\pm$0.03 & 16.92$\pm$0.05 \\
19.9 & 57571.0 & NOT & 16.34$\pm$0.13 & 17.16$\pm$0.03 & -- & 17.07$\pm$0.03 & 17.06$\pm$0.05 & 16.94$\pm$0.05 \\
22.5 & 57573.9 & \emph{pt5m} & -- & -- & -- & -- & 17.04$\pm$0.02 & 17.17$\pm$0.06 \\
23.4 & 57574.9 & \emph{pt5m} & -- & -- & -- & -- & 17.04$\pm$0.02 & 17.06$\pm$0.08 \\
29.7 & 57581.8 & \emph{Gaia} & -- & -- & -- & 17.40$\pm$0.07 & -- & -- \\
29.7 & 57581.9 & \emph{Gaia} & -- & -- & -- & 17.37$\pm$0.06 & -- & -- \\
38.0 & 57590.9 & NOT & -- & 18.35$\pm$0.08 & -- & 17.46$\pm$0.07 & 17.45$\pm$0.06 & 17.25$\pm$0.06 \\
38.2 & 57591.2 & \textit{Swift} & 17.77$\pm$0.18 & 18.52$\pm$0.24 & -- & 17.71$\pm$0.25 & -- & -- \\
41.7 & 57595.0 & \textit{Swift} & 17.64$\pm$0.20 & 18.03$\pm$0.20 & -- & 17.83$\pm$0.26 & -- & -- \\
130.8 & 57693.2 & NOT & -- & -- & -- & *19.48$\pm$0.08 & *19.68$\pm$0.10 & *19.47$\pm$0.09 \\
138.3 & 57701.5 & \emph{Gaia} & -- & -- & -- & *19.83$\pm$0.08 & -- & -- \\
162.4 & 57728.1 & NOT & 21.07$\pm$0.18 & *22.04$\pm$0.15 & -- & *20.24$\pm$0.13 & *20.35$\pm$0.10 & *20.12$\pm$0.09 \\
169.8 & 57736.2 & NOT & -- & *22.28$\pm$0.15 & -- & *20.44$\pm$0.13 & *20.77$\pm$0.10 & *20.72$\pm$0.09 \\
183.4 & 57751.2 & NOT & -- & *22.30$\pm$0.18 & -- & *20.66$\pm$0.17 & *20.79$\pm$0.10 & *20.96$\pm$0.09 \\
209.7 & 57780.2 & NOT & -- & *23.83$\pm$0.16 & -- & *21.49$\pm$0.13 & *21.85$\pm$0.10 & *22.17$\pm$0.09 \\
\hline
\end{tabular}
\label{table:phot}
\end{small}
\end{minipage}
\end{table*}

\begin{table}
\begin{small}
\caption{Log of \emph{Swift} UV photometric observations of Gaia16apd. Magnitudes are reported in the Vega system. No $K$-corrections have been applied.}
\centering
\begin{tabular}{rlccc}
\hline
Phase & MJD & $uvw2$ & $uvm2$ & $uvw1$ \\
(d) &  & (mag) & (mag) & (mag) \\
\hline
$-17.6$ & 57529.7 & 15.71$\pm$0.04 & 15.27$\pm$0.04 & 15.15$\pm$0.04 \\
$-15.8$ & 57531.7 & 15.76$\pm$0.04 & 15.26$\pm$0.04 & 15.19$\pm$0.05 \\
$-14.5$ & 57533.2 & 15.77$\pm$0.04 & -- & 15.14$\pm$0.05 \\
$-10.6$ & 57537.5 & 15.82$\pm$0.04 & 15.37$\pm$0.04 & 15.16$\pm$0.04 \\
$-6.9$ & 57541.5 & 15.91$\pm$0.04 & 15.40$\pm$0.04 & 15.16$\pm$0.04 \\
$-5.2$ & 57543.3 & 15.97$\pm$0.05 & 15.53$\pm$0.05 & 15.24$\pm$0.05 \\
$-3.5$ & 57545.3 & 16.15$\pm$0.04 & 15.66$\pm$0.04 & 15.36$\pm$0.05 \\
$-1.6$ & 57547.3 & 16.11$\pm$0.05 & 15.67$\pm$0.04 & 15.41$\pm$0.04 \\
1.3 & 57550.6 & 16.24$\pm$0.05 & 15.82$\pm$0.04 & 15.47$\pm$0.04 \\
7.1 & 57556.9 & 16.63$\pm$0.05 & 16.22$\pm$0.05 & 15.95$\pm$0.05 \\
9.3 & 57559.4 & 16.78$\pm$0.05 & 16.46$\pm$0.05 & -- \\
11.8 & 57562.1 & 17.18$\pm$0.05 & 16.62$\pm$0.05 & 16.25$\pm$0.06 \\
38.2 & 57591.2 & 19.31$\pm$0.20 & 19.26$\pm$0.17 & 18.86$\pm$0.26 \\
41.7 & 57595.0 & 19.08$\pm$0.26 & 19.40$\pm$0.31 & 18.50$\pm$0.28 \\
 \hline
\end{tabular}
\label{table:phot2}
\end{small}
\end{table}

\begin{table}
\begin{small}
\caption{Log of NOT/NOTCam NIR photometric observations of Gaia16apd. Magnitudes are reported in the Vega system. No $K$-corrections have been applied.}
\centering
\begin{tabular}{rlccc}
\hline
Phase & MJD & $J$ & $H$ & $K$ \\
(d) &  & (mag) & (mag) & (mag) \\
\hline
1.7 & 57551.0 & 16.59$\pm$0.08 & 16.23$\pm$0.15 & 16.59$\pm$0.15 \\
11.6 & 57561.9 & 16.32$\pm$0.08 & 16.79$\pm$0.15 & 16.85$\pm$0.15 \\
31.6 & 57583.9 & 16.20$\pm$0.08 & 16.68$\pm$0.15 & 16.35$\pm$0.15 \\
57.0 & 57611.9 & 17.13$\pm$0.08 & 17.36$\pm$0.15 & 17.25$\pm$0.15 \\
141.7 & 57705.2 & 18.51$\pm$0.08 & 18.49$\pm$0.15 & 17.94$\pm$0.15 \\
171.6 & 57738.2 & 18.79$\pm$0.08 & 18.69$\pm$0.15 & -- \\
194.3 & 57763.2 & 19.54$\pm$0.08 & 19.42$\pm$0.15 & 19.03$\pm$0.15 \\
 \hline
\end{tabular}
\label{table:phot3}
\end{small}
\end{table}

\begin{table}
\begin{small}
\centering
\caption{$K$-corrections applied in different bands close to the $g$-band peak, obtained using the ALFOSC optical spectrum at 3.5 d and the \emph{HST}/STIS UV spectrum \citep{yan16} at 3.7 d. The $K$-corrections are \emph{added} to the uncorrected magnitudes.}
\begin{tabular}{cccc}
\hline
Observed band & Corrected band & $K$-correction \\
 & & (mag) \\
\hline
$uvw2$ (Vega) & $uvw2$ (Vega) & -0.23$\pm$0.10\\
$uvm2$ (Vega) & $uvm2$ (Vega) & -0.24$\pm$0.01\\
$uvw1$ (Vega) & $uvw1$ (Vega) & -0.17$\pm$0.12\\
$U$ (Vega) & $U$ (Vega) & -0.01$\pm$0.02\\
$B$ (Vega) & $B$ (Vega) & 0.18$\pm$0.01\\
$V$ (Vega) & $g$ (AB) & 0.08$\pm$0.01\\
$g$ (AB) & $g$ (AB) & 0.18$\pm$0.01\\
$r$ (AB) & $r$ (AB) & 0.21$\pm$0.01\\
$i$ (AB) & $i$ (AB) & 0.30$\pm$0.01\\
 \hline
\end{tabular}
\label{table:kcorr}
\end{small}
\end{table}

\begin{figure*}
\begin{minipage}{1.\linewidth}
\centering
\includegraphics[width=0.8\columnwidth]{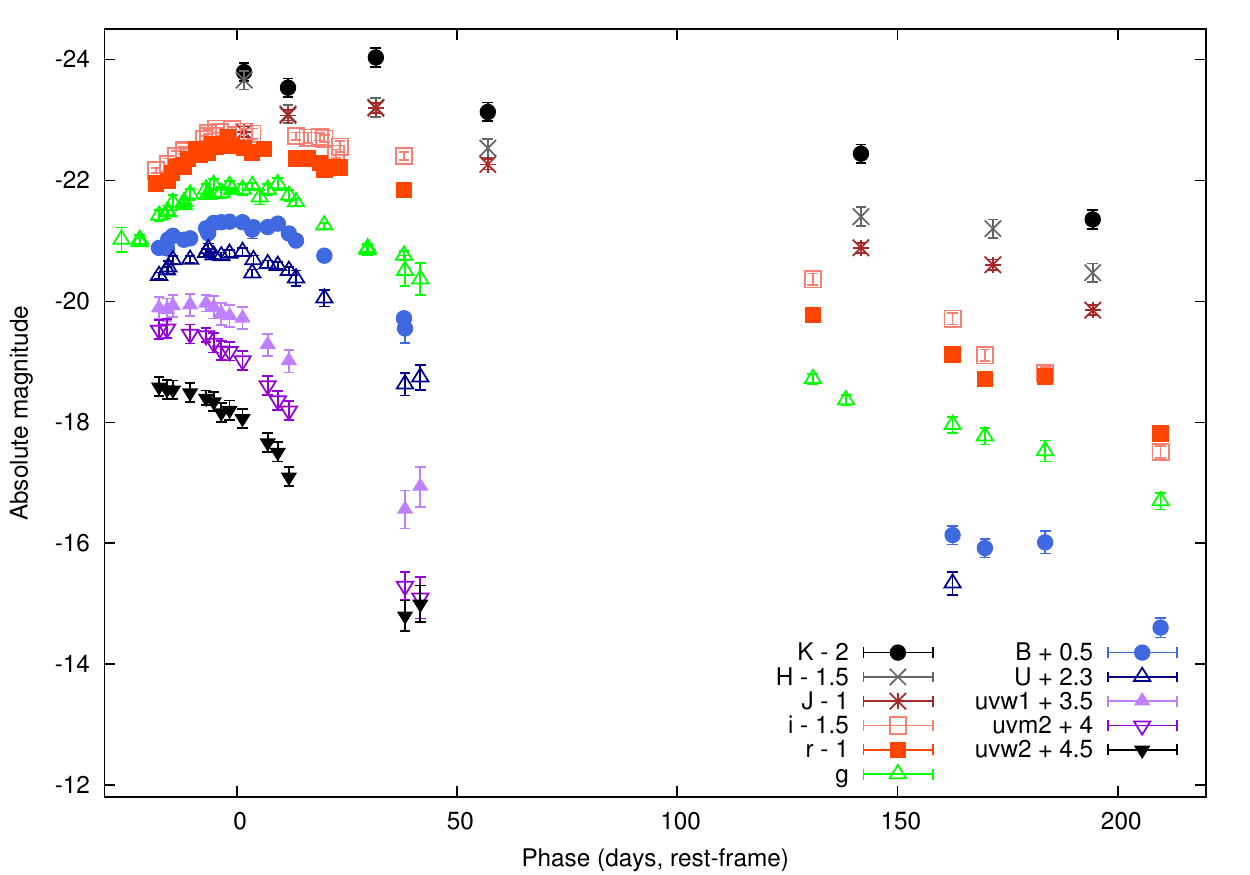}
\caption{Extinction- and $K$-corrected absolute magnitude light curves of Gaia16apd. The reference date is the $g$-band peak.}
\label{fig:uvoir_lc}
\end{minipage}
\end{figure*}

\begin{figure}
\centering
\includegraphics[width=\columnwidth]{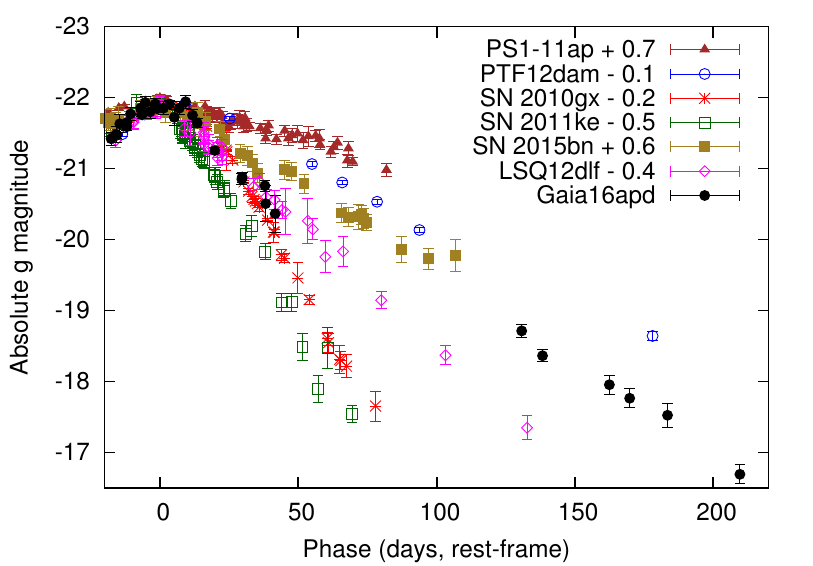} \\
\includegraphics[width=\columnwidth]{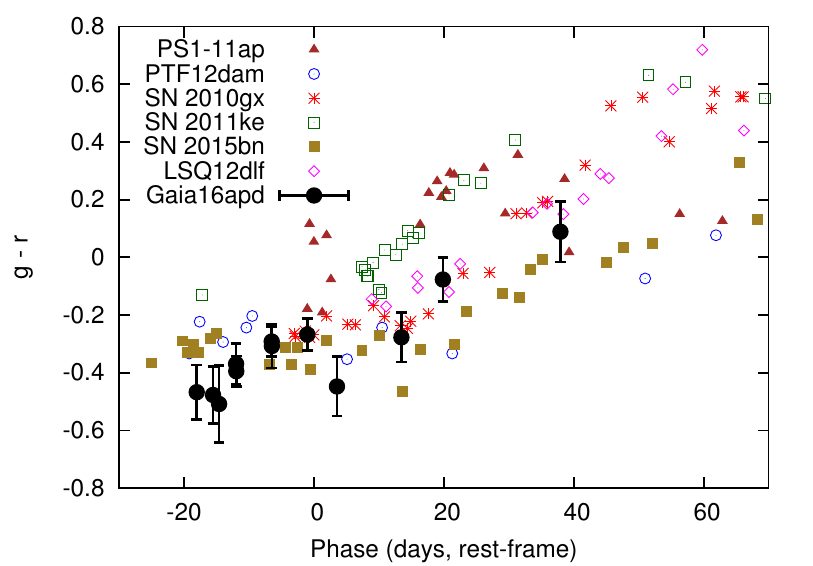} \\
\includegraphics[width=\columnwidth]{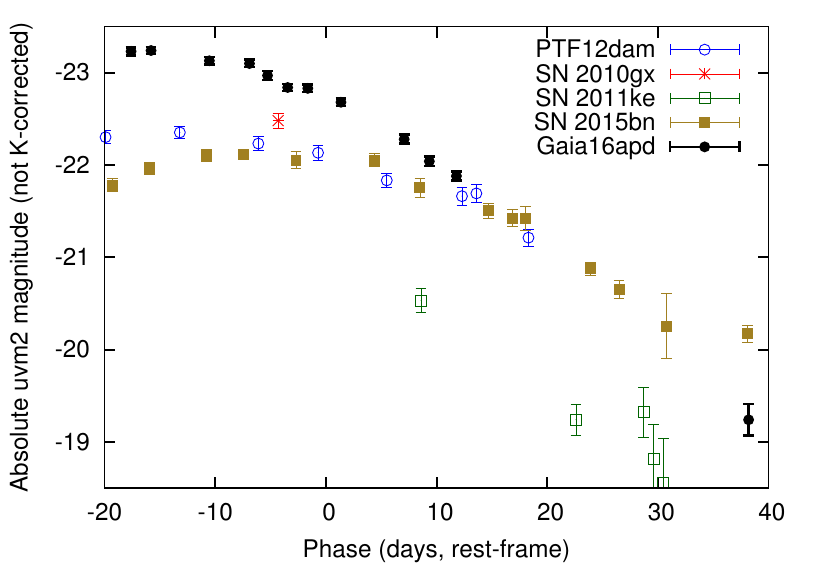}
\caption{Extinction- and $K$-corrected absolute $g$-band light curve (top; shifted to approximately the same peak magnitude), corrected $g-r$ colour evolution (middle) and non-corrected UV light curve (if available; bottom) of Gaia16apd and six comparison events: PS1-11ap \citep[dark red;][]{mccrum14}, PTF12dam \citep[blue;][]{nicholl13}, SN~2010gx \citep[red;][]{pastorello10}, SN~2011ke \citep[green;][]{inserra13}, SN~2015bn \citep[light brown;][]{nicholl16} and LSQ12dlf \citep[magenta;][]{nicholl14}. The $K$-corrections of the comparison events were done using SNAKE and publicly available spectra. The errors of the $g-r$ colours of the comparison events have been omitted for clarity. The reference date is the $g$-band peak.}
\label{fig:lc_comp}
\end{figure}

The early photometry has not been corrected for host galaxy contamination. However, given that the host galaxy is faint, $g = 21.73 \pm 0.06$ mag (over 4 mag fainter than the SN around peak), this contamination is on the order of a few per cent around the peak and has no significant effect on our analysis. At 130 d, the contamination is on the order of 15 per cent, and the host galaxy magnitudes (from SDSS) have been subtracted from the late-time photometry, using the \citet{jordi06} conversions when needed. The photometry was corrected for Galactic extinction before $K$-correction, and for host extinction (see Section \ref{sec:host}) after $K$-correction, using the \citet{cardelli89} law. SDSS magnitudes are reported in the AB system, while other magnitudes, including UV, are in the Vega system. A complete log of photometric observations is presented in Tables \ref{table:phot} (optical data), \ref{table:phot2} (UV data) and  \ref{table:phot3} (NIR data). The \emph{Swift} magnitudes before MJD 57591 have also been reported by \citet{yan16}. We note that although the \emph{Swift} $uvw1$ and $uvw2$ filters suffer from `red leak', i.e. contamination by redder photons \citep{brown10}, for objects as blue as SLSNe this is not significant (Brown, private communication).

The optical photometry was $K$-corrected using the SuperNova Algorithm for $K$-correction Evaluation (SNAKE) code \citep{inserra16} based on optical spectra of Gaia16apd. $K$-corrections in the $U$ band were determined partially using blackbody fits to the spectra. \emph{Swift} UV bands were $K$-corrected using \emph{HST}/STIS spectra published by \citet{yan16}. NIR bands were not $K$-corrected, as corrections using optical spectra were deemed unreliable. For intervening epochs, $K$-corrections were obtained through polynomial interpolation. The $V$-band magnitudes were $K$-corrected to the $g$ band for clarity and to facilitate comparisons to other SLSNe. The $K$-corrections around the $g$-band peak are listed in Table \ref{table:kcorr}.

The absolute magnitudes in all observed bands (after $K$- and extinction correction) are presented in Figure \ref{fig:uvoir_lc}. In Figure \ref{fig:lc_comp}, we also show a comparison between Gaia16apd and six spectroscopically similar SLSNe -- PS1-11ap \citep{mccrum14}, PTF12dam \citep{nicholl13}, SN~2010gx \citep{pastorello10}, SN~2011ke \citep{inserra13}, SN~2015bn \citep{nicholl16} and LSQ12dlf \citep{nicholl14} -- in $g$, the $g-r$ colour evolution, and the \emph{Swift} $uvm2$ band. 

Using a low-order polynomial fit to the absolute $g$-band light curve, we obtained the peak epoch MJD = 57549.1 $\pm$ 0.9 d and peak magnitude $M_{g,\mathrm{peak}} = -21.8 \pm 0.1$ mag. As shown in Figure \ref{fig:lc_comp}, in the $g$-band Gaia16apd declines faster at early times than the slowly-declining PTF12dam and similarly to SNe 2010gx and 2011ke, while being optically slightly brighter at peak. However, it then settles into a linear decline of 1.9 mag (100 d)$^{-1}$ at $\sim$40 d, and the decline after this point is similar to PTF12dam and SN~2015bn, making Gaia16apd photometrically intermediate between the fast- and slowly-declining events. LSQ12dlf evolves in a similar fashion: it initially resembles SNe 2010gx and 2011ke, but its decline slows down after $\sim50$ d, although not becoming quite as slow as PTF12dam or SN~2015bn during the observations. Furthermore, the $r$-band decline of both Gaia16apd and LSQ12dlf -- less than 2 mag (100 d)$^{-1}$ -- is similar to slowly-declining events from the beginning. Thus LSQ12dlf seems to be another photometrically intermediate event and, together with Gaia16apd, suggests a continuum of light-curve properties between the fast- and slowly-declining type Ic SLSNe. Gaia16apd starts off slightly bluer in $g-r$ than SN~2010gx, evolving from roughly $-0.5$ mag at $-15$ d to $\sim$0.1 mag at 40 d. From $\sim10$ days after the maximum their colour evolution, and that of LSQ12dlf, is similar. SN~2011ke remains $\sim0.3$ mag redder than Gaia16apd, while the slow type Ic SLSNe are intially slightly redder but redden more slowly.

Gaia16apd reaches an absolute magnitude of $M_{uvm2} = -23.55 \pm 0.22$ mag (Vega). However, as host extinction correction and $K$-correction in the UV have not been done in earlier SLSN studies, we here compare them to the non-corrected absolute magnitude of $M_{uvm2} = -23.25 \pm 0.05$ mag. As the UV magnitudes of the comparison SNe are not $K$-corrected, we include $uvw1$ in the following comparison for the more distant events, where the observed $uvw1$ is closer to $uvm2$ as observed at the redshift of Gaia16apd, $z=0.102$. Gaia16apd is initially roughly a magnitude brighter in the UV than either PTF12dam or SN~2015bn, another slowly-declining SLSN with good UV coverage -- PTF12dam (at $z = 0.107$) reaches $M_{uvm2} = -22.35 \pm 0.08$ mag and SN~2015bn ($z = 0.1136$) reaches $M_{uvm2} = -22.11 \pm 0.05$ mag. However, eventually Gaia16apd declines to a brightness similar to them at 10 -- 20 d. Out of the fast-declining events of \citet{inserra13}, PTF11rks ($z = 0.190$) reaches $M_{uvm2} = -20.80 \pm 0.13$ mag and $M_{uvw1} = -21.21 \pm 0.12$ mag -- other SLSNe in the sample have no early \emph{Swift} observations, but are between 0.3 and 1.3 mag fainter than Gaia16apd at 25 -- 30 d. One early epoch of \emph{Swift} data exists for SN~2010gx ($z = 0.23$), yielding $M_{uvm2} = -22.48 \pm 0.08$ mag and $M_{uvw1} = -22.87 \pm 0.06$ mag at $-5.6$ d, respectively 0.5 mag and 0.1 mag fainter than Gaia16apd at this epoch. In the NIR, Gaia16apd and PTF12dam reach a non-$K$-corrected $M_{H} = -22.16 \pm 0.16$ and $-22.17 \pm 0.07$ mag (Vega), respectively, around $g$-band peak, but PTF12dam again declines slower afterwards.

\subsection{Spectroscopy} \label{sec:spec}

Optical long-slit spectroscopic follow-up observations were performed using NOT/ALFOSC and the Optical System for Imaging and low-Intermediate-Resolution Integrated Spectroscopy (OSIRIS) instrument on the 10.4-m Gran Telescopio Canarias (GTC). The reduction of the NOT/ALFOSC spectroscopy was done using {\sc foscgui} or the {\sc quba} pipeline, and the reduction of the GTC/OSIRIS spectroscopy was done using a custom pipeline running standard {\sc iraf} tasks. Cosmic rays were removed using {\sc lacosmic} \citep{lacosmic}. Sensitivity curves were obtained using spectroscopic standard stars observed during the same night. A complete log of spectroscopic observations is presented in Table \ref{table:speclog}.

\begin{table*}
\begin{minipage}{1.\linewidth}
\begin{small}
\caption{Log of spectroscopic observations of Gaia16apd.}
\centering
\begin{tabular}{rllccccc}
\hline
 Phase\footnote{Rest-frame days since $g$-band peak at MJD 57549.1.} & MJD & Date & Instrument & Wavelength & Slit width & Resolution & Exposure   \\ 
   (d) & 	&  	&	& (\AA)	&	(arcsec)	& (km/s) &	(s)	     \\ \hline
 \hline
$-18.3$ & 57528.9 & 2016 May 20.9 & NOT/ALFOSC & 3200 -- 9600 & 1.0 & 700 & 1800  \\
$-15.5$ & 57532.0 & 2016 May 24.0 & NOT/ALFOSC & 3200 -- 9600 & 1.0 & 700 & 1800  \\
3.5 & 57553.0 & 2016 June 14.0 & NOT/ALFOSC & 3200 -- 9600 & 1.3 & 900 & 1500  \\
13.5 & 57564.0 & 2016 June 25.0 & NOT/ALFOSC & 3200 -- 9600 & 1.0 & 700 & 1500  \\
22.5 & 57573.9 & 2016 July 4.9 & NOT/ALFOSC & 3200 -- 9600 & 1.0 & 700 & 1500  \\
32.5 & 57584.9 & 2016 July 15.9 & NOT/ALFOSC & 3200 -- 9600 & 1.3 & 900 & 1500  \\
44.3 & 57597.9 & 2016 July 28.9 & NOT/ALFOSC & 3200 -- 9600 & 1.0 & 700 & 1800  \\
151.6 & 57716.2 & 2016 November 24.2 & NOT/ALFOSC & 3200 -- 9600 & 1.0 & 700 & 3600 \\
166.2 & 57732.2 & 2016 December 10.2 & NOT/ALFOSC & 3200 -- 9600 & 1.0 & 700 & 3600 \\
197.9 & 57767.2 & 2017 January 14.2 & GTC/OSIRIS & 3600 -- 7200 & 1.0 & 550 & 3600 \\
197.9 & 57767.2 & 2017 January 14.2 & GTC/OSIRIS & 4800 -- 10000 & 1.0 & 500 & 1800 \\
234.1 & 57807.1 & 2017 February 23.1 & GTC/OSIRIS & 4800 -- 10000 & 1.0 & 500 & 3600 \\
 \hline
\end{tabular}
\label{table:speclog}
\end{small}
\end{minipage}
\end{table*}

\begin{figure*}
\begin{minipage}{1.\linewidth}
\centering
\includegraphics[width=0.7\columnwidth]{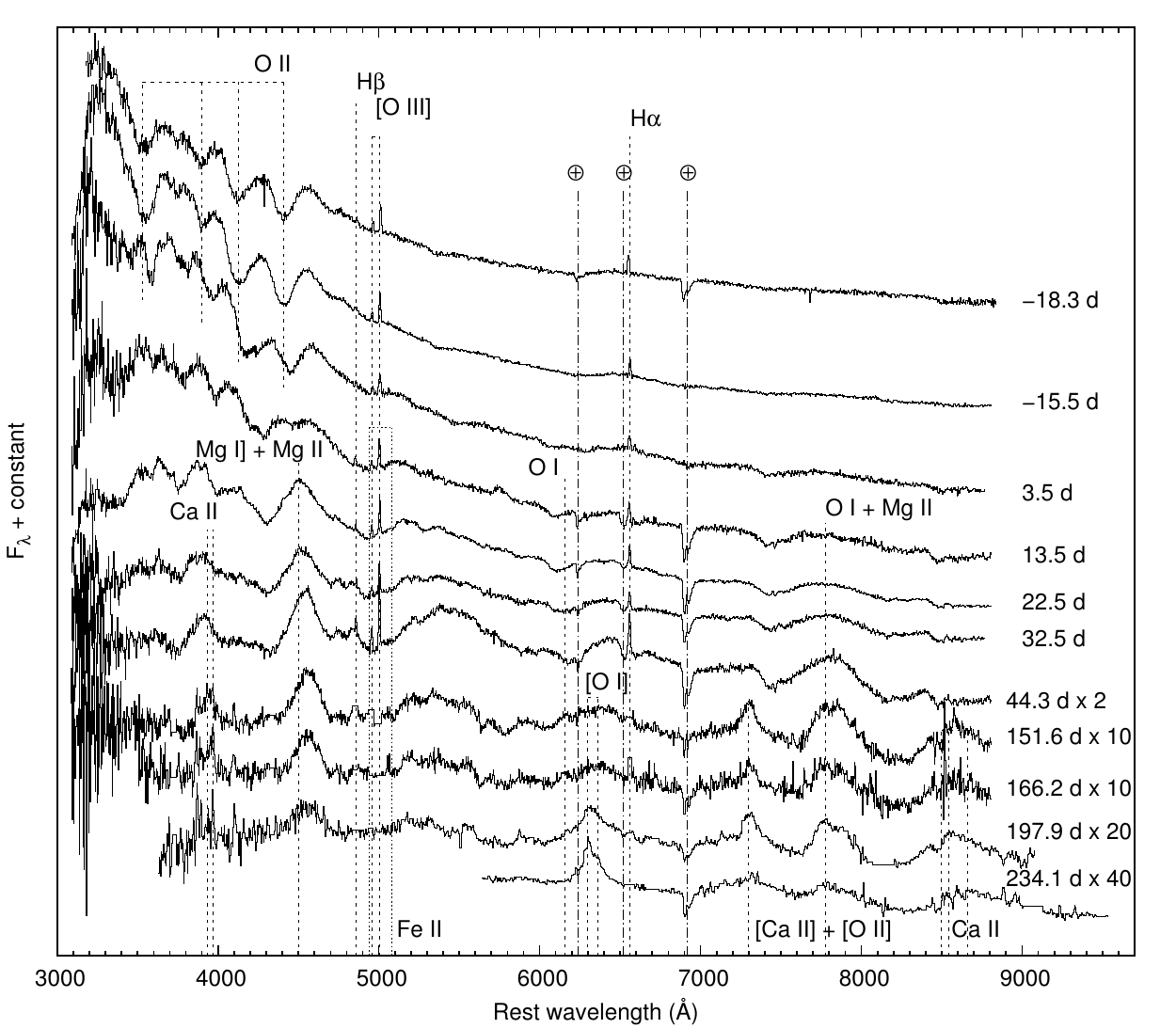} \\
\includegraphics[width=0.7\columnwidth]{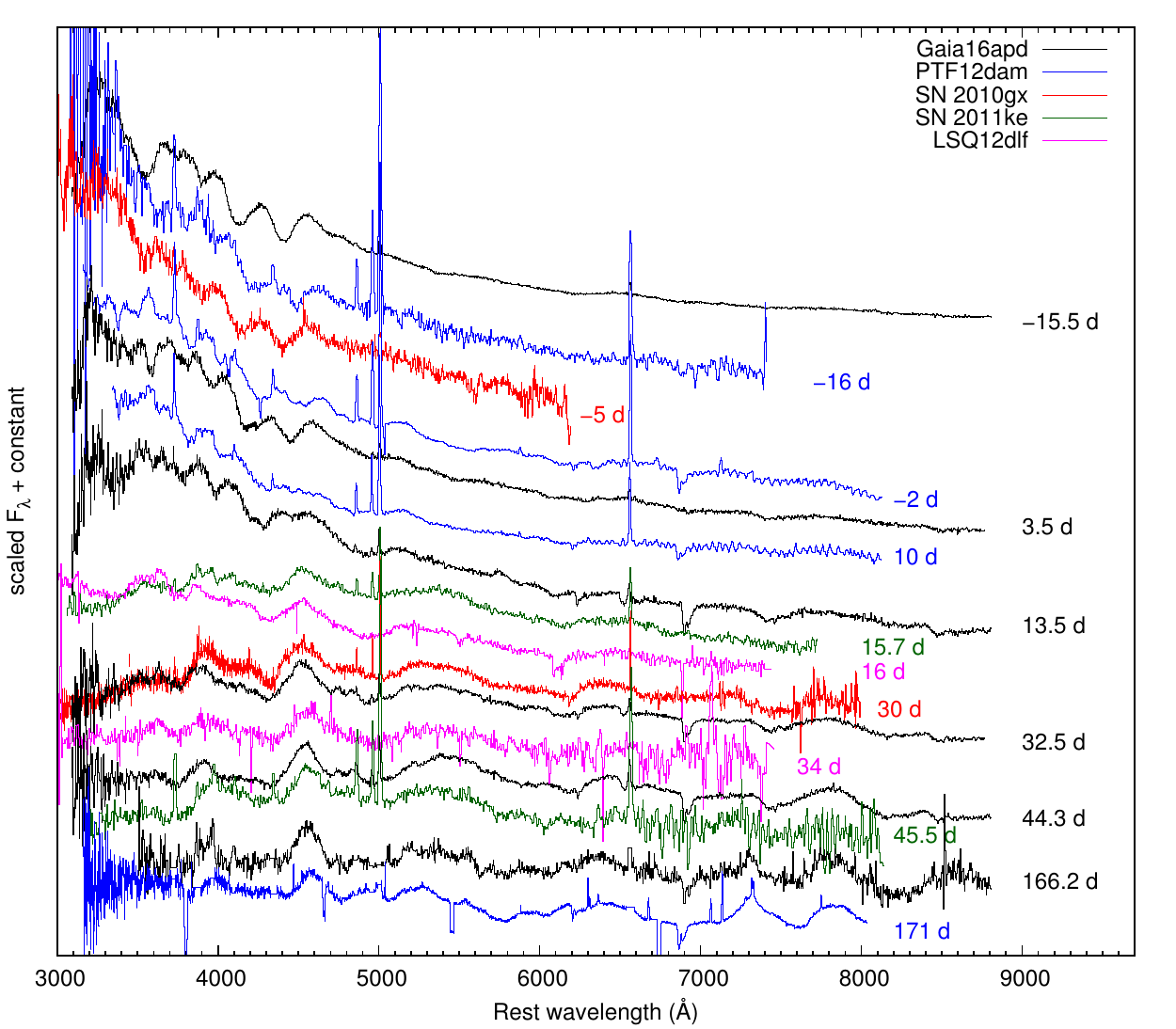} 
\caption{Spectral evolution of Gaia16apd (top) and a comparison to other type Ic SLSNe, SN~2010gx \citep{pastorello10}, PTF12dam \citep{nicholl13}, SN~2011ke \citep{inserra13} and LSQ12dlf \citep{nicholl14} (bottom). The Gaia16apd spectra show prominent O {\sc ii} absorption lines at early stages, typical for type Ic SLSNe. Strong, narrow host galaxy lines have been removed from late-time ($>150$ d) spectra for clarity. The reference date is the $g$-band peak after $K$-correction for Gaia16apd, SN~2011ke and LSQ12dlf, the $r$-band peak for PTF12dam and the $B$-band peak for SN~2010gx.}
\label{fig:specseq}
\end{minipage}
\end{figure*}

The spectra are presented in Figure \ref{fig:specseq}, along with a comparison to PTF12dam, SN~2010gx, SN~2011ke and LSQ12dlf. The spectra from $-15.5$ d to 3.5 d are characterized by the distinctive O {\sc ii} absorption lines around 3500 -- 4500 \AA, typical for type Ic SLSNe \citep{quimby11}, with gradually decreasing velocity -- the minimum of the $\lambda\lambda4415,4417$ absorption changes from $\sim-$19800 km s$^{-1}$ to $\sim-$15600 km s$^{-1}$. No other clear features were identified on top of the hot continua of the pre-peak spectra. By day 3.5, a weak Fe {\sc ii} $\lambda5169$ P Cygni profile is visible. By day 22.5, the O {\sc ii} lines have been replaced by Fe {\sc ii} absorption lines, Mg {\sc ii} $\lambda4481$ absorption and Mg {\sc i}] $\lambda4571$ emission, and the spectra resemble normal type Ic SNe around maximum light; this transition is still ongoing on day 13.5. O {\sc i} $\lambda$6156 and $\lambda$7774 absorption is also visible, along with a Ca {\sc ii} $\lambda\lambda3969,3750$ P Cygni profile and Mg {\sc ii} $\lambda\lambda7877,7896$ emission. The photospheric velocity, measured from the minima of the Fe {\sc ii} $\lambda5169$ absorption profiles, stays nearly constant between days 3.5 and 44.3, evolving from $\sim12700$ km s$^{-1}$ to $\sim12400$ km s$^{-1}$. Such a nearly flat velocity evolution is common among both fast- and slowly-declining type Ic SLSNe \citep{nicholl15}.

\begin{figure*}
\begin{minipage}{1.\linewidth}
\centering
\includegraphics[width=1\columnwidth]{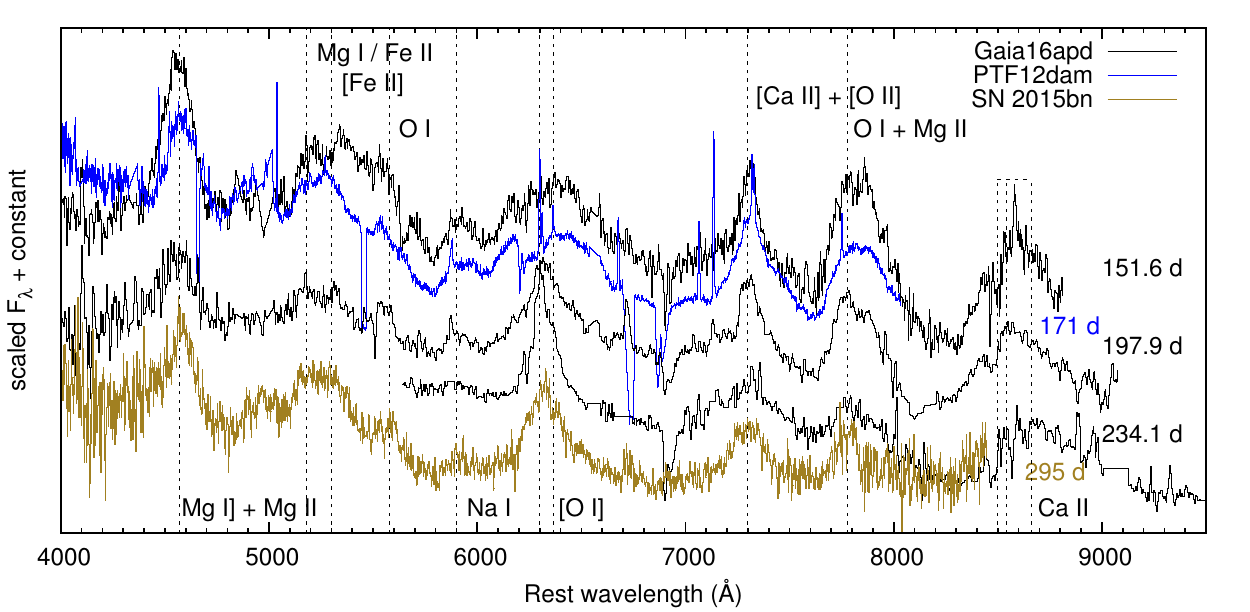}  
\caption{Comparison of late-time spectra of Gaia16apd (black), PTF12dam \citep[blue;][]{nicholl13} and SN~2015bn \citep[light brown;][]{nicholl16}, demonstrating their similarity. Line identifications are based on \citet{nicholl13} and \citet{jerkstrand16}. Strong, narrow host galaxy lines have been removed for clarity.}
\label{fig:spec_late}
\end{minipage}
\end{figure*}

The features described above are common to type Ic SLSNe; as seen in the bottom panel of Figure \ref{fig:specseq}, our comparison SNe show a similar evolution. The O {\sc ii} absorption lines in SN~2010gx disappear somewhat faster, indicating a lower degree of ionization (temperature), as in the other events in \citet{inserra13}, where O {\sc ii} is weak or nonexistent a few days after the $g$-band peak. The slower-declining PTF12dam also only shows the O {\sc ii} lines until a few days before maximum light. The later evolution ($\gtrsim 15$ d) was very similar in all these events. 

Figure \ref{fig:spec_late} shows a clearer comparison between the late-time spectra of Gaia16apd (at 151.6, 197.9 and 234.1 d) and the slowly declining SLSNe PTF12dam (at 171 d) and SN~2015bn (at 295 d). The continuum visible in the late-time spectra is attributed partially to host galaxy contamination, since at 197.9 d the SN is approaching the host galaxy magnitudes. However, as \citet{inserra17} pointed out, such a continuum is often present in the late-time spectra of slowly-declining type Ic SLSNe, possibly due to interaction with a small amount of CSM. Thus the spectra could be called `pseudo-nebular'. The spectrum of Gaia16apd at these phases is dominated by broad emission lines of calcium ([Ca {\sc ii}] $\lambda\lambda$7291,7324 and the NIR triplet), oxygen ([O {\sc i}] $\lambda\lambda$6300,6364 and O {\sc i} $\lambda$7776), sodium, magnesium and iron, at typical FWHM velocities of 10000 km s$^{-1}$. The exception to this is the [Ca {\sc ii}]/[O {\sc ii}] feature at $\sim7300$~\AA, with a width of $\sim4000$ km s$^{-1}$, superposed on another emission feature similar to O {\sc i} $\lambda$7776 in width. The narrower peak, visible in the spectra of slowly-declining events, suggests multiple emitting regions contributing to the spectra; \citet{inserra17} attributed it to emission from the inner, slower-moving and more diffuse regions of the ejecta. The narrower [Ca {\sc ii}]/[O {\sc ii}] peak is replaced by a strong [O {\sc i}] $\lambda\lambda$6300,6364 peak with a width of $\sim6000$ km s$^{-1}$ between 166.2 and 234.1 d. A similar evolution takes place in PTF12dam and SN~2015bn as well, but at a later date \citep[after $\sim250$ d;][]{chen15,nicholl16c}. Apart from the earlier appearance of the strong [O {\sc i}] $\lambda\lambda$6300,6364 peak, the late-time spectra of Gaia16apd, PTF12dam and SN~2015bn are strikingly similar, despite their different early photometric evolution. The similar line ratios imply similar temperatures and elemental abundances. Unfortunately, no spectra at such a late phase exist for the other photometrically intermediate event LSQ12dlf (or for SN~2010gx-like fast events), but at 106 d it does exhibit similar broad emission lines as Gaia16apd at 151.6 d \citep{nicholl14}.

\section{Analysis} \label{sec:analysis}

\subsection{SED and bolometric light curve} \label{sec:sed}

$K$- and extinction-corrected magnitudes were converted to fluxes to construct a spectral enegy distribution (SED) for each epoch. Polynomial interpolation was used to obtain fluxes in all bandpasses for epochs where some filters were not used. The resulting SEDs were found to be in agreement with the UV and optical spectra. Pseudo-bolometric luminosities were calculated by integrating the SED over the bands from $uvw2$ to $K$, assuming zero flux outside these filters. A low-order polynomial fit to the pseudo-bolometric light curve yielded a peak epoch of MJD = 57539.6 $\pm$ 1.1 d and a peak luminosity of 2.9 $\pm$ 0.1 $\times$ 10$^{44}$ erg~s$^{-1}$.

Blackbody functions were fitted to the fluxes using the {\sc mpfit} routine \citep{mpfit} in IDL. The evolution of the blackbody temperature, pseudo-bolometric luminosity and $L_{griz}$ (see below) are presented in Figure \ref{fig:bbpar}. The temperature evolved smoothly from $\sim$19000 K at $-18$ d to $\sim$8000 K at 30 d. The estimated blackbody temperatures of other type I SLSNe are similar after the $g$-band peak, around 12000 -- 14000 K a few days after the peak, but somewhat cooler than Gaia16apd before it \citep[e.g.][]{inserra13}. This is consistent with how the $g-r$ colour evolution and UV brightness of Gaia16apd compare to the events in Figure \ref{fig:lc_comp}.

For the purpose of comparison with the \citet{nicholl15} sample, we have also calculated $L_{griz}$, the luminosity over the $griz$ filters. The $z$-band fluxes were estimated by integrating the SDSS $z$-band filter over the blackbody function at each epoch. These were then combined with the $gri$ fluxes to construct an SED, which was integrated over wavelength assuming zero flux outside the $griz$ filters, following \citet{inserra13}. This $griz$-bolometric luminosity was expressed in the form of a magnitude, adapting the standard formula for the bolometric magnitude:
\begin{equation}
M_{griz} = -2.5 \mathrm{log_{10}} \frac{L_{griz}}{3.055 \times 10^{35}\mathrm{erg~s^{-1}}} .
\end{equation}

\begin{figure*}
\begin{minipage}{1.\linewidth}
\centering
\includegraphics[width=0.8\columnwidth]{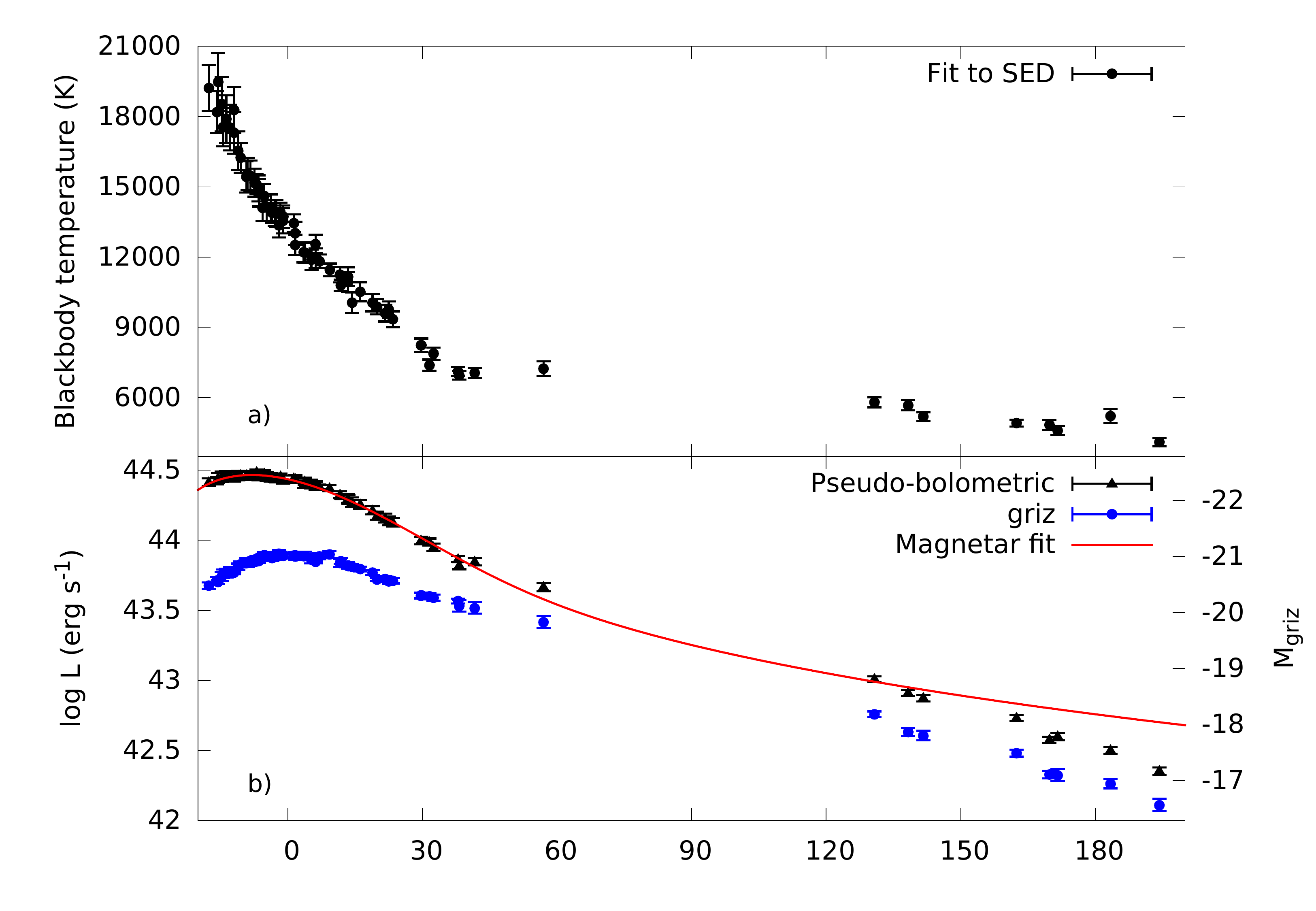} 
\caption{Evolution of the blackbody temperature (a) and the pseudo-bolometric (black) and $griz$-band (blue) luminosity (b). The red line in (b) is a magnetar model fit with magnetic field $1.9 \times 10^{14}$ G, period 1.9 ms and diffusion time 41 d.}
\label{fig:bbpar}
\end{minipage}
\end{figure*}

Using a low-order polynomial fit, the $griz$ luminosity peak epoch was estimated as MJD = 57549.2 $\pm$ 0.4 d and the peak $M_{griz}$ as $-21.0 \pm$ 0.1 mag, somewhat brighter than most fast-declining type Ic SLSNe but fainter than SN~2005ap at $-21.22$ \citep{nicholl15}. From this fit, we also estimated the $e$-folding rise time $\tau_{\mathrm{rise}}$ of the $griz$-band luminosity as 23 d. The decline time $\tau_{\mathrm{dec}}$ was estimated as 47 d. Both values must be considered rough estimates, as the multi-band photometry at our disposal does not cover the phases before -18 d or between 42 and 130 d. The $\tau_{\mathrm{dec}}$ is between the typical values for fast- and slowly-declining events. The values are, however, consistent with the empirical $\tau_{\mathrm{dec}} \approx 2\tau_{\mathrm{rise}}$ relation \citep{nicholl15}.

\subsection{Modelling} \label{sec:model}

The injection of rotational energy from the rapid spin-down of a newborn highly magnetized neutron star (magnetar), with a magnetic field $B$ in a range of $10^{14}$ -- $10^{15}$ G and an initial spin period $P$ of a few ms, can significantly boost the optical luminosity of a SN \citep{kasenbildsten10,woosley10}. Millisecond magnetar models have been successfully employed to reproduce the light curves of type I SLSNe \citep[e.g.][]{chomiuk11,inserra13,lunnan16}. $P$ determines the rotational energy $E_p \simeq 2 \times 10^{52}$ erg $\times$ ($P$/1 ms)$^{-2}$, while $B$ and $P$ together determine the spin-down time-scale $\tau_p \simeq 4.7$ d $\times (P$/1 ms)$^{2} \times (B$/10$^{14}$ G)$^{-2}$. The third input parameter, the diffusion time in the ejecta $\tau_m$, is determined by the opacity $\kappa$, ejecta mass $M_{\mathrm{ej}}$ and kinetic energy $E_{\mathrm{k}}$ as 
\begin{equation}
\tau_m = 10 \mathrm{d} \left(\frac{M_{\mathrm{ej}}}{1~\mathrm{M}_{\odot}}\right)^{3/4} \left(\frac{E_{\mathrm{k}}}{10^{51}~\mathrm{erg}}\right)^{-1/4} \left(\frac{\kappa}{0.1~\mathrm{cm}^2~\mathrm{g}^{-1}}\right)^{1/2}.
\end{equation}
The luminosity evolution $L(t)$ of the SLSN, assuming it to be dominated by the magnetar power, is then described by
\begin{equation}
L(t) = \frac{2E_p}{\tau_p \tau_m} e^{-(\frac{t}{\tau_m})^2} \int_{0}^{t} \frac{1}{(1 + t'/\tau_p)^2} e^{(\frac{t'}{\tau_m})^2} \frac{t'dt'}{\tau_m} .
\end{equation}

%This model was fitted to the pseudo-bolometric light curve of Gaia16apd using Markov Chain Monte Carlo (MCMC) simulations based on the {\sc python} package {\sc emcee} \citep{emcee}. We used 40,000 iterations in total, while discarding the initial 10,000 as part of the burn-in sample. The best values were derived using the maximum of the posterior probability distribution, and the error bars correspond to 1 $\sigma$ integrated probability towards lower and upper values of the distribution. 

This model was fitted to the pseudo-bolometric light curve of Gaia16apd until 42 d, applying a grid of input parameters at intervals of 0.05 ms, 0.05 $\times$ 10$^{14}$ G and 1 d for $P$, $B$ and $\tau_m$ respectively, for a total of 8000 fits. The final parameters were estimated as the average and standard deviation of the fits with $\chi^{2}_{\nu} < 2$. The resulting parameters were $P = 1.9\pm0.2$ ms, $B = 1.9\pm0.2 \times 10^{14}$ G and $\tau_m = 41\pm3$ d. The derived explosion date was MJD = 57512 (from a rest-frame rise time of 28 d). The best-fitting magnetar light curve is shown in Figure \ref{fig:bbpar}. This fit no longer provides a good match to the linear decline rate at late times ($>140$ d), overestimating the flux by a factor of $\sim4.5$ by 210 d. This may be due to less efficient trapping of the magnetar energy at these times. Approximating the rise of the bolometric light curve using the magnetar fit, the total radiated energy by day 210 is $1.6 \times 10^{51}$ erg. Using Eq. (5) of \citet{nicholl15}, we estimated the ejecta mass as
\begin{equation}
M_{\mathrm{ej}} = 7.7 \times 10^{-7}~\mathrm{M}_{\odot} \times \left(\frac{\kappa}{0.1~\mathrm{cm}^2~\mathrm{g}^{-1}}\right)^{-1} \frac{v}{\mathrm{km~s}^{-1}} \left(\frac{\tau_m}{\mathrm{d}}\right)^2 ,
\end{equation}
where $v$ is approximately the photospheric velocity, measured using the minima of Fe {\sc ii} $\lambda5169$ absorption lines. As the velocity evolution of this line is nearly flat, we used the average velocity between 3.5 and 44.3 d, 12500 km s$^{-1}$. Assuming $\kappa = 0.1~\mathrm{cm}^2~\mathrm{g}^{-1}$ \citep[shown to be a reasonable assumption for electron-scattering-dominated opacity by][]{inserra13}, the resulting estimated ejecta mass is $\sim16~\mathrm{M}_\odot$ (with $E_{k} \simeq 2.5\times10^{52}$ erg), while with $\kappa = 0.2~\mathrm{cm}^2~\mathrm{g}^{-1}$ (full ionization), the result is $\sim 8~\mathrm{M}_\odot$ ($E_{k} \simeq 1.3\times10^{52}$ erg) -- the latter estimate is close to the ejecta masses of some abnormal type Ic SNe \citep{valenti12,taddia16}. A high degree of ionization could be caused by hard radiation from the magnetar \citep{metzger14}.

The kinetic energy has been estimated assuming that the entire mass of the ejecta is swept up into a thin shell and moves at the same velocity, which is a reasonable approximation in the magnetar scenario \citep{kasenbildsten10}. \citet{nicholl16b} arrived at a much lower value of $2.4 \times 10^{51}$ erg despite very similar magnetar parameters -- which they acknowledged as a probable underestimate, as they assumed homologous expansion. The ejecta mass and kinetic energy estimated by \citet{yan16} ($12~M_\odot$ and $>2 \times 10^{52}$ erg, respectively) are closer to ours.

\section{Discussion and conclusions} \label{sec:disc}

We have presented observations of Gaia16apd, one of the closest SLSNe ever discovered \citep[$z = 0.102$; the only SLSN reported at a similar distance was PTF10hgi at $z=0.100$;][]{inserra13}, in a wavelength range spanning the \emph{Swift} UV bands, the optical region and NIR. Spectroscopically Gaia16apd is a typical type Ic SLSN; photometrically, it is intermediate between fast-declining type Ic SLSNe such as SN~2011ke and slowly-declining ones such as PTF12dam. While its peak absolute magnitudes of $M_g = -21.8 \pm 0.1$ and $M_{griz} = -21.0 \pm 0.1$ are bright for the fast-declining class, with a $B$-band peak of $\sim-21.8$ and a decline of $\sim1.1$ mag in the first 30 d after peak it is close to the \citet{inserra14} peak-decline relation for SN~2005ap-like events, i.e. fast-declining type Ic SLSNe. We do note that \citet{inserra14} used a synthetic bandpass around 4000~\AA~and not the $B$ band. The host galaxy is a faint, low-metallicity dwarf galaxy consistent with those of other type Ic SLSNe \citep{perley16}.

The nearly constant Fe {\sc ii} $\lambda5169$ line velocity is consistent with a magnetar-powered explosion \citep{kasenbildsten10}, as the central engine is expected to sweep up most of the ejecta into a shell instead of the normal homologous expansion scenario. A dense shell can, however, also be created in the scenarios powered by CSM interaction \citep[e.g.][]{cf94} or a black hole central engine \citep{dexterkasen13}. \citet{yan16} found a lack of line blanketing in the UV spectra of Gaia16apd compared to normal SNe, caused by a lack of iron-group elements such as $^{56}$Ni in the ejecta, and concluded that a PISN scenario was implausible based on the low $^{56}$Ni mass \citep[although, as shown by][the line blanketing is not significantly weaker than in other SLSNe]{nicholl16b}. 

We have thus fitted a magnetar model to the pseudo-bolometric light curve and obtained best-fit parameters of $P = 1.9\pm0.2$ ms, $B = 1.9\pm0.2 \times 10^{14}$ G and $\tau_m = 41\pm3$ d. These values are consistent with those predicted for newborn magnetars based on theory \citep{dt92}. The total radiated energy until 210 d is $1.6 \times 10^{51}$ erg, on the order of 10 per cent of the estimated kinetic energy. The rise time to the bolometric peak was estimated as 28 d. The ejecta mass obtained using $\tau_m$ and the Fe {\sc ii} $\lambda5169$ velocity is 8 or 16 $\mathrm{M}_\odot$, with opacities $\kappa = 0.2$ \citep[possible for magnetar-powered explosions;][]{metzger14} or $0.1~\mathrm{cm}^2~\mathrm{g}^{-1}$, respectively. The ejecta mass and decline time-scale are between the typical values for fast- and slowly-declining type Ic SLSNe, but Gaia16apd does conform to a $\tau_{\mathrm{dec}} \approx 2\tau_{\mathrm{rise}}$ relation. This also suggests a magnetar origin, as such a tight relation is difficult to explain using CSM interaction models \citep[by e.g.][]{chatz12} without fine-tuning the mass and density of the CSM \citep{nicholl15}. However, the magnetar model no longer provides a good fit to the light curve after $\sim140$ d, possibly indicating a lessening opacity and trapping of the magnetar energy over time. \citet{nicholl16b} obtained a good fit to their light curve, but without late-time photometry after 140 d. Figure \ref{fig:bbpar} shows that the data up to this point do provide a reasonable fit, explaining the apparent discrepancy.

Gaia16apd exhibited UV magnitudes significantly brighter than those of slow-declining type Ic SLSNe with UV coverage. Of the fast-declining events, only SN~2010gx showed a similar early UV brightness. The pre-peak temperature of Gaia16apd was somewhat higher than usual for its class, consistently with the early colour evolution and UV brightness. The only other superluminous event with a similar temperature and UV-optical colour is the nuclear transient ASASSN-15lh \citep{dong16}, but this transient was more likely a tidal disruption event than an actual SLSN \citep{leloudas16}. \citet{yan16} argued that the lack of iron-group elements in the outer ejecta is the main reason for the UV brightness, while \citet{nicholl16b} showed that it can simply be explained with the right combination of magnetar parameters. With our magnetar fit, the central engine power \citep[which determines the colour of the spectrum as shown in Figure 11 of][]{howell13} is consistent with theirs, and we agree with the latter interpretation. This event highlights the importance of good rest-frame UV coverage. Apart from SN~2010gx and Gaia16apd, other similarly UV-bright examples are not known so far, possibly because of insufficient UV observations. SLSNe have been proposed as standard candles observable at high redshift \citep{inserra14,scova16}, which will require a robust description and understanding of their UV variability. 

The late-time spectra at 151.6 and 166.2 d are very similar to that of PTF12dam at 171 d, with several strong, broad emission lines of oxygen, magnesium and calcium, and a narrower emission feature of [Ca {\sc ii}]/[O {\sc ii}]. The latter was attributed by \citet{inserra17} to emission from the inner regions, i.e. from inside the cavity created by the pulsar wind, and was considered indirect evidence of a central engine scenario. The earlier appearance of the strong [O {\sc i}] $\lambda\lambda$6300,6364 in Gaia16apd compared to slowly-declining events reflects a faster temperature evolution in the inner cavity and thus the shorter spin-down time-scale of the magnetar. Spectra of SLSNe photometrically more similar to events such as SN~2010gx at $>100$ d are needed to assess how their late-time evolution fits this picture, but the faster evolution is consistent with a continuum between fast and slow type Ic SLSNe. Together with the photometric evolution of Gaia16apd and of LSQ12dlf, intermediate between the fast- and slowly-declining type Ic SLSNe, all this suggests that the two subclasses form a continuum of properties instead of having discrete progenitor populations and mechanisms.

\citet{jerkstrand16} applied nebular-phase models to the late-time spectra of slowly-declining type Ic SLSNe and showed them to be consistent with $\gtrsim 10~\mathrm{M}_{\odot}$ of oxygen-rich ejecta, which places a limit of $\gtrsim 40~\mathrm{M}_{\odot}$ on the Wolf-Rayet progenitor in a single-star scenario. The similarity between Gaia16apd, PTF12dam and SN~2015bn, and by proxy other slowly-declining type Ic SLSNe, suggests that this may also hold true for at least some faster-declining events. For Gaia16apd, $M_{\mathrm{ej}} \gtrsim 10~\mathrm{M}_{\odot}$ is also consistent with our estimate from the magnetar model. PISN and CSM interaction-dominated models do not seem to match observations of either subclass \citep[e.g.][]{nicholl14,mccrum14,yan16,lunnan16}, and a millisecond magnetar scenario is now widely considered to be the most likely one for both. However, \citet{inserra17} find the slowly-declining events consistent with a magnetar engine combined with interaction with a small amount of CSM. In addition to ejecta mass, the strength of the interaction may be one of the factors influencing the photometric evolution of type Ic SLSNe.

\section*{Acknowledgements}

We thank the anonymous referee for comments that helped improve this paper considerably. We thank Subo Dong, Cosimo Inserra, Christa Gall, Peter Brown and Andrea Pastorello for their suggestions.

We acknowledge ESA Gaia, DPAC and the Photometric Science Alerts Team (http://gsaweb.ast.cam.ac.uk/alerts). The \emph{pt5m} and TJO preliminary photometric calibrations were obtained using the Cambridge Photometric Calibration Server (CPCS), designed and maintained by Sergey Koposov and Lukasz Wyrzykowski. 

Based on observations made with the Nordic Optical Telescope, operated by the Nordic Optical Telescope Scientific Association, and with the Gran Telescopio Canarias (GTC), at the Observatorio del Roque de los Muchachos, La Palma, Spain, of the Instituto de Astrofisica de Canarias. The data presented here were obtained in part with ALFOSC, which is provided by the Instituto de Astrofisica de Andalucia (IAA) under a joint agreement with the University of Copenhagen and NOTSA. We also want to thank support astronomers working in Telescopi Joan Or\'{o} at Montsec Observatory (OAdM) for their help to run the needed sequences on time for our Gaia alert programme.

T.~K. acknowledges financial support by the Emil Aaltonen Foundation. N.~B. was supported by the GROWTH project funded by the National Science Foundation under Grant No 1545949. J. H. acknowledges support by the Finnish Cultural Foundation. M.~F. acknowledges the support of a Royal Society - Science Foundation Ireland University Research Fellowship. N.~E.~R. acknowledges financial support by the 1994 PRIN-INAF 2014 (project `Transient Universe: unveiling new types of stellar explosions with PESSTO') and by MIUR PRIN 2010- 2011, ``The dark Universe and the cosmic evolution of baryons: from current surveys to Euclid". E.~Y.~H. acknowledges the support provided by the National Science Foundation under Grant No. AST-1008343 and AST-1613472. M.~D.~S. is funded by generous support provided by the Danish Agency for Science and Technology and Innovation realised through a Sapere Aude Level 2 grant, and a grant from the Villum Foundation. This work was supported by the MINECO (Spanish Ministry of Economy) - FEDER through grants ESP2016-80079-C2-1-R and ESP2014-55996-C2-1-R and MDM-2014-0369 of ICCUB (Unidad de Excelencia 'Mar\'{i}a de Maeztu'). NUTS is funded in part by the IDA (Instrument Centre for Danish Astronomy). \L{}.~W. was supported by Polish National Science Centre grant OPUS 2015/17/B/ST9/03167.

%%%%%%%%%%%%%%%%%%%%%%%%%%%%%%%%%%%%%%%%%%%%%%%%%%



\bibliographystyle{mnras}
 \bibliography{kangas16apd} % if your bibtex file is called example.bib

\begin{thebibliography}{}
\makeatletter
\relax
\def\mn@urlcharsother{\let\do\@makeother \do\$\do\&\do\#\do\^\do\_\do\%\do\~}
\def\mn@doi{\begingroup\mn@urlcharsother \@ifnextchar [ {\mn@doi@}
  {\mn@doi@[]}}
\def\mn@doi@[#1]#2{\def\@tempa{#1}\ifx\@tempa\@empty \href
  {http://dx.doi.org/#2} {doi:#2}\else \href {http://dx.doi.org/#2} {#1}\fi
  \endgroup}
\def\mn@eprint#1#2{\mn@eprint@#1:#2::\@nil}
\def\mn@eprint@arXiv#1{\href {http://arxiv.org/abs/#1} {{\tt arXiv:#1}}}
\def\mn@eprint@dblp#1{\href {http://dblp.uni-trier.de/rec/bibtex/#1.xml}
  {dblp:#1}}
\def\mn@eprint@#1:#2:#3:#4\@nil{\def\@tempa {#1}\def\@tempb {#2}\def\@tempc
  {#3}\ifx \@tempc \@empty \let \@tempc \@tempb \let \@tempb \@tempa \fi \ifx
  \@tempb \@empty \def\@tempb {arXiv}\fi \@ifundefined
  {mn@eprint@\@tempb}{\@tempb:\@tempc}{\expandafter \expandafter \csname
  mn@eprint@\@tempb\endcsname \expandafter{\@tempc}}}

\bibitem[\protect\citeauthoryear{{Bennett}, {Larson}, {Weiland}  \&
  {Hinshaw}}{{Bennett} et~al.}{2014}]{cosmology}
{Bennett} C.~L.,  {Larson} D.,  {Weiland} J.~L.,   {Hinshaw} G.,  2014, \mn@doi
  [\apj] {10.1088/0004-637X/794/2/135}, \href
  {http://adsabs.harvard.edu/abs/2014ApJ...794..135B} {794, 135}

\bibitem[\protect\citeauthoryear{{Bertin} \& {Arnouts}}{{Bertin} \&
  {Arnouts}}{1996}]{bertin96}
{Bertin} E.,  {Arnouts} S.,  1996, \mn@doi [\aaps] {10.1051/aas:1996164}, \href
  {http://adsabs.harvard.edu/abs/1996A%26AS..117..393B} {117, 393}

\bibitem[\protect\citeauthoryear{{Blagorodnova}, {Yan}, {Quimby}, {Olaes},
  {Brown}  \& {Cooke}}{{Blagorodnova} et~al.}{2016}]{swift_atel}
{Blagorodnova} N.,  {Yan} L.,  {Quimby} R.,  {Olaes} M.~K.,  {Brown} P.,
  {Cooke} J.,  2016, The Astronomer's Telegram, \href
  {http://adsabs.harvard.edu/abs/2016ATel.9074....1B} {9074}

\bibitem[\protect\citeauthoryear{{Brown} et~al.,}{{Brown}
  et~al.}{2010}]{brown10}
{Brown} P.~J.,  et~al., 2010, \mn@doi [\apj] {10.1088/0004-637X/721/2/1608},
  \href {http://adsabs.harvard.edu/abs/2010ApJ...721.1608B} {721, 1608}

\bibitem[\protect\citeauthoryear{{Cardelli}, {Clayton}  \& {Mathis}}{{Cardelli}
  et~al.}{1989}]{cardelli89}
{Cardelli} J.~A.,  {Clayton} G.~C.,   {Mathis} J.~S.,  1989, \mn@doi [\apj]
  {10.1086/167900}, \href {http://adsabs.harvard.edu/abs/1989ApJ...345..245C}
  {345, 245}

\bibitem[\protect\citeauthoryear{{Chatzopoulos}, {Wheeler}  \&
  {Vinko}}{{Chatzopoulos} et~al.}{2012}]{chatz12}
{Chatzopoulos} E.,  {Wheeler} J.~C.,   {Vinko} J.,  2012, \mn@doi [\apj]
  {10.1088/0004-637X/746/2/121}, \href
  {http://adsabs.harvard.edu/abs/2012ApJ...746..121C} {746, 121}

\bibitem[\protect\citeauthoryear{{Chen} et~al.,}{{Chen} et~al.}{2013}]{chen13}
{Chen} T.-W.,  et~al., 2013, \mn@doi [\apjl] {10.1088/2041-8205/763/2/L28},
  \href {http://adsabs.harvard.edu/abs/2013ApJ...763L..28C} {763, L28}

\bibitem[\protect\citeauthoryear{{Chen} et~al.,}{{Chen} et~al.}{2015}]{chen15}
{Chen} T.-W.,  et~al., 2015, \mn@doi [\mnras] {10.1093/mnras/stv1360}, \href
  {http://adsabs.harvard.edu/abs/2015MNRAS.452.1567C} {452, 1567}

\bibitem[\protect\citeauthoryear{{Chevalier} \& {Fransson}}{{Chevalier} \&
  {Fransson}}{1994}]{cf94}
{Chevalier} R.~A.,  {Fransson} C.,  1994, \mn@doi [\apj] {10.1086/173557},
  \href {http://adsabs.harvard.edu/abs/1994ApJ...420..268C} {420, 268}

\bibitem[\protect\citeauthoryear{{Chevalier} \& {Irwin}}{{Chevalier} \&
  {Irwin}}{2011}]{chevalierirwin11}
{Chevalier} R.~A.,  {Irwin} C.~M.,  2011, \mn@doi [\apjl]
  {10.1088/2041-8205/729/1/L6}, \href
  {http://adsabs.harvard.edu/abs/2011ApJ...729L...6C} {729, L6}

\bibitem[\protect\citeauthoryear{{Chomiuk} et~al.,}{{Chomiuk}
  et~al.}{2011}]{chomiuk11}
{Chomiuk} L.,  et~al., 2011, \mn@doi [\apj] {10.1088/0004-637X/743/2/114},
  \href {http://adsabs.harvard.edu/abs/2011ApJ...743..114C} {743, 114}

\bibitem[\protect\citeauthoryear{{Craig} et~al.,}{{Craig}
  et~al.}{2015}]{craig15}
{Craig} M.~W.,  et~al., 2015, {ccdproc: CCD data reduction software},
  Astrophysics Source Code Library (\mn@eprint {ascl} {1510.007})

\bibitem[\protect\citeauthoryear{{Dessart}, {Waldman}, {Livne}, {Hillier}  \&
  {Blondin}}{{Dessart} et~al.}{2013}]{dessart13}
{Dessart} L.,  {Waldman} R.,  {Livne} E.,  {Hillier} D.~J.,   {Blondin} S.,
  2013, \mn@doi [\mnras] {10.1093/mnras/sts269}, \href
  {http://adsabs.harvard.edu/abs/2013MNRAS.428.3227D} {428, 3227}

\bibitem[\protect\citeauthoryear{{Dexter} \& {Kasen}}{{Dexter} \&
  {Kasen}}{2013}]{dexterkasen13}
{Dexter} J.,  {Kasen} D.,  2013, \mn@doi [\apj] {10.1088/0004-637X/772/1/30},
  \href {http://adsabs.harvard.edu/abs/2013ApJ...772...30D} {772, 30}

\bibitem[\protect\citeauthoryear{{Djupvik} \& {Andersen}}{{Djupvik} \&
  {Andersen}}{2010}]{notref}
{Djupvik} A.~A.,  {Andersen} J.,  2010, \mn@doi [Astrophysics and Space Science
  Proceedings] {10.1007/978-3-642-11250-8_21}, \href
  {http://adsabs.harvard.edu/abs/2010ASSP...14..211D} {14, 211}

\bibitem[\protect\citeauthoryear{{Dong} et~al.,}{{Dong} et~al.}{2016}]{dong16}
{Dong} S.,  et~al., 2016, \mn@doi [Science] {10.1126/science.aac9613}, \href
  {http://adsabs.harvard.edu/abs/2016Sci...351..257D} {351, 257}

\bibitem[\protect\citeauthoryear{{Duncan} \& {Thompson}}{{Duncan} \&
  {Thompson}}{1992}]{dt92}
{Duncan} R.~C.,  {Thompson} C.,  1992, \mn@doi [\apjl] {10.1086/186413}, \href
  {http://adsabs.harvard.edu/abs/1992ApJ...392L...9D} {392, L9}

\bibitem[\protect\citeauthoryear{{Gaia Collaboration} et~al.,}{{Gaia
  Collaboration} et~al.}{2016}]{gaia}
{Gaia Collaboration} et~al., 2016, \mn@doi [\aap]
  {10.1051/0004-6361/201629272}, \href
  {http://adsabs.harvard.edu/abs/2016A%26A...595A...1G} {595, A1}

\bibitem[\protect\citeauthoryear{{Gal-Yam}}{{Gal-Yam}}{2012}]{galyam12}
{Gal-Yam} A.,  2012, \mn@doi [Science] {10.1126/science.1203601}, \href
  {http://adsabs.harvard.edu/abs/2012Sci...337..927G} {337, 927}

\bibitem[\protect\citeauthoryear{{Gal-Yam} et~al.,}{{Gal-Yam}
  et~al.}{2009}]{galyam10}
{Gal-Yam} A.,  et~al., 2009, \mn@doi [\nat] {10.1038/nature08579}, \href
  {http://adsabs.harvard.edu/abs/2009Natur.462..624G} {462, 624}

\bibitem[\protect\citeauthoryear{{G{\"u}ver} \& {{\"O}zel}}{{G{\"u}ver} \&
  {{\"O}zel}}{2009}]{guver09}
{G{\"u}ver} T.,  {{\"O}zel} F.,  2009, \mn@doi [\mnras]
  {10.1111/j.1365-2966.2009.15598.x}, \href
  {http://adsabs.harvard.edu/abs/2009MNRAS.400.2050G} {400, 2050}

\bibitem[\protect\citeauthoryear{{Hardy}, {Butterley}, {Dhillon}, {Littlefair}
  \& {Wilson}}{{Hardy} et~al.}{2015}]{pt5m}
{Hardy} L.~K.,  {Butterley} T.,  {Dhillon} V.~S.,  {Littlefair} S.~P.,
  {Wilson} R.~W.,  2015, \mn@doi [\mnras] {10.1093/mnras/stv2279}, \href
  {http://adsabs.harvard.edu/abs/2015MNRAS.454.4316H} {454, 4316}

\bibitem[\protect\citeauthoryear{{Heger} \& {Woosley}}{{Heger} \&
  {Woosley}}{2002}]{hegerwoosley02}
{Heger} A.,  {Woosley} S.~E.,  2002, \mn@doi [\apj] {10.1086/338487}, \href
  {http://adsabs.harvard.edu/abs/2002ApJ...567..532H} {567, 532}

\bibitem[\protect\citeauthoryear{{Howell} et~al.,}{{Howell}
  et~al.}{2013}]{howell13}
{Howell} D.~A.,  et~al., 2013, \mn@doi [\apj] {10.1088/0004-637X/779/2/98},
  \href {http://adsabs.harvard.edu/abs/2013ApJ...779...98H} {779, 98}

\bibitem[\protect\citeauthoryear{{Inserra} \& {Smartt}}{{Inserra} \&
  {Smartt}}{2014}]{inserra14}
{Inserra} C.,  {Smartt} S.~J.,  2014, \mn@doi [\apj]
  {10.1088/0004-637X/796/2/87}, \href
  {http://adsabs.harvard.edu/abs/2014ApJ...796...87I} {796, 87}

\bibitem[\protect\citeauthoryear{{Inserra} et~al.,}{{Inserra}
  et~al.}{2013}]{inserra13}
{Inserra} C.,  et~al., 2013, \mn@doi [\apj] {10.1088/0004-637X/770/2/128},
  \href {http://adsabs.harvard.edu/abs/2013ApJ...770..128I} {770, 128}

\bibitem[\protect\citeauthoryear{{Inserra} et~al.,}{{Inserra}
  et~al.}{2016}]{inserra16}
{Inserra} C.,  et~al., 2016, preprint, \href
  {http://adsabs.harvard.edu/abs/2016arXiv160401226I} {} (\mn@eprint {arXiv}
  {1604.01226})

\bibitem[\protect\citeauthoryear{{Inserra} et~al.,}{{Inserra}
  et~al.}{2017}]{inserra17}
{Inserra} C.,  et~al., 2017, preprint, \href
  {http://adsabs.harvard.edu/abs/2017arXiv170100941I} {} (\mn@eprint {arXiv}
  {1701.00941})

\bibitem[\protect\citeauthoryear{{Jerkstrand} et~al.,}{{Jerkstrand}
  et~al.}{2017}]{jerkstrand16}
{Jerkstrand} A.,  et~al., 2017, \mn@doi [\apj] {10.3847/1538-4357/835/1/13},
  \href {http://adsabs.harvard.edu/abs/2017ApJ...835...13J} {835, 13}

\bibitem[\protect\citeauthoryear{{Jordi}, {Grebel}  \& {Ammon}}{{Jordi}
  et~al.}{2006}]{jordi06}
{Jordi} K.,  {Grebel} E.~K.,   {Ammon} K.,  2006, \mn@doi [\aap]
  {10.1051/0004-6361:20066082}, \href
  {http://adsabs.harvard.edu/abs/2006A%26A...460..339J} {460, 339}

\bibitem[\protect\citeauthoryear{{Jordi} et~al.,}{{Jordi}
  et~al.}{2010}]{jordi10}
{Jordi} C.,  et~al., 2010, \mn@doi [\aap] {10.1051/0004-6361/201015441}, \href
  {http://adsabs.harvard.edu/abs/2010A%26A...523A..48J} {523, A48}

\bibitem[\protect\citeauthoryear{{Kangas} et~al.,}{{Kangas}
  et~al.}{2016}]{classif}
{Kangas} T.,  et~al., 2016, The Astronomer's Telegram, \href
  {http://adsabs.harvard.edu/abs/2016ATel.9071....1K} {9071}

\bibitem[\protect\citeauthoryear{{Kasen} \& {Bildsten}}{{Kasen} \&
  {Bildsten}}{2010}]{kasenbildsten10}
{Kasen} D.,  {Bildsten} L.,  2010, \mn@doi [\apj]
  {10.1088/0004-637X/717/1/245}, \href
  {http://adsabs.harvard.edu/abs/2010ApJ...717..245K} {717, 245}

\bibitem[\protect\citeauthoryear{{Kasen}, {Woosley}  \& {Heger}}{{Kasen}
  et~al.}{2011}]{kasen11}
{Kasen} D.,  {Woosley} S.~E.,   {Heger} A.,  2011, \mn@doi [\apj]
  {10.1088/0004-637X/734/2/102}, \href
  {http://adsabs.harvard.edu/abs/2011ApJ...734..102K} {734, 102}

\bibitem[\protect\citeauthoryear{{Kobulnicky}, {Kennicutt}  \&
  {Pizagno}}{{Kobulnicky} et~al.}{1999}]{kobulnicky99}
{Kobulnicky} H.~A.,  {Kennicutt} Jr. R.~C.,   {Pizagno} J.~L.,  1999, \mn@doi
  [\apj] {10.1086/306987}, \href
  {http://adsabs.harvard.edu/abs/1999ApJ...514..544K} {514, 544}

\bibitem[\protect\citeauthoryear{{Lang}, {Hogg}, {Mierle}, {Blanton}  \&
  {Roweis}}{{Lang} et~al.}{2010}]{lang10}
{Lang} D.,  {Hogg} D.~W.,  {Mierle} K.,  {Blanton} M.,   {Roweis} S.,  2010,
  \mn@doi [\aj] {10.1088/0004-6256/139/5/1782}, \href
  {http://adsabs.harvard.edu/abs/2010AJ....139.1782L} {139, 1782}

\bibitem[\protect\citeauthoryear{{Leloudas} et~al.,}{{Leloudas}
  et~al.}{2015}]{leloudas15}
{Leloudas} G.,  et~al., 2015, \mn@doi [\mnras] {10.1093/mnras/stv320}, \href
  {http://adsabs.harvard.edu/abs/2015MNRAS.449..917L} {449, 917}

\bibitem[\protect\citeauthoryear{{Leloudas} et~al.,}{{Leloudas}
  et~al.}{2016}]{leloudas16}
{Leloudas} G.,  et~al., 2016, \mn@doi [Nature Astronomy]
  {10.1038/s41550-016-0002}, \href
  {http://adsabs.harvard.edu/abs/2016NatAs...1E...2L} {1, 0002}

\bibitem[\protect\citeauthoryear{{Lunnan} et~al.,}{{Lunnan}
  et~al.}{2015}]{lunnan15}
{Lunnan} R.,  et~al., 2015, \mn@doi [\apj] {10.1088/0004-637X/804/2/90}, \href
  {http://adsabs.harvard.edu/abs/2015ApJ...804...90L} {804, 90}

\bibitem[\protect\citeauthoryear{{Lunnan} et~al.,}{{Lunnan}
  et~al.}{2016}]{lunnan16}
{Lunnan} R.,  et~al., 2016, \mn@doi [\apj] {10.3847/0004-637X/831/2/144}, \href
  {http://adsabs.harvard.edu/abs/2016ApJ...831..144L} {831, 144}

\bibitem[\protect\citeauthoryear{{Markwardt}}{{Markwardt}}{2009}]{mpfit}
{Markwardt} C.~B.,  2009, in {Bohlender} D.~A.,  {Durand} D.,   {Dowler} P.,
  eds,  Astronomical Society of the Pacific Conference Series Vol. 411,
  Astronomical Data Analysis Software and Systems XVIII. p.~251 (\mn@eprint
  {arXiv} {0902.2850})

\bibitem[\protect\citeauthoryear{{Mattila} et~al.,}{{Mattila}
  et~al.}{2016}]{nuts}
{Mattila} S.,  et~al., 2016, The Astronomer's Telegram, \href
  {http://adsabs.harvard.edu/abs/2016ATel.8992....1M} {8992}

\bibitem[\protect\citeauthoryear{{McCrum} et~al.,}{{McCrum}
  et~al.}{2014}]{mccrum14}
{McCrum} M.,  et~al., 2014, \mn@doi [\mnras] {10.1093/mnras/stt1923}, \href
  {http://adsabs.harvard.edu/abs/2014MNRAS.437..656M} {437, 656}

\bibitem[\protect\citeauthoryear{{McCrum} et~al.,}{{McCrum}
  et~al.}{2015}]{mccrum15}
{McCrum} M.,  et~al., 2015, \mn@doi [\mnras] {10.1093/mnras/stv034}, \href
  {http://adsabs.harvard.edu/abs/2015MNRAS.448.1206M} {448, 1206}

\bibitem[\protect\citeauthoryear{{Metzger}, {Vurm}, {Hasco{\"e}t}  \&
  {Beloborodov}}{{Metzger} et~al.}{2014}]{metzger14}
{Metzger} B.~D.,  {Vurm} I.,  {Hasco{\"e}t} R.,   {Beloborodov} A.~M.,  2014,
  \mn@doi [\mnras] {10.1093/mnras/stt1922}, \href
  {http://adsabs.harvard.edu/abs/2014MNRAS.437..703M} {437, 703}

\bibitem[\protect\citeauthoryear{{Nicholl} et~al.,}{{Nicholl}
  et~al.}{2013}]{nicholl13}
{Nicholl} M.,  et~al., 2013, \mn@doi [\nat] {10.1038/nature12569}, \href
  {http://adsabs.harvard.edu/abs/2013Natur.502..346N} {502, 346}

\bibitem[\protect\citeauthoryear{{Nicholl} et~al.,}{{Nicholl}
  et~al.}{2014}]{nicholl14}
{Nicholl} M.,  et~al., 2014, \mn@doi [\mnras] {10.1093/mnras/stu1579}, \href
  {http://adsabs.harvard.edu/abs/2014MNRAS.444.2096N} {444, 2096}

\bibitem[\protect\citeauthoryear{{Nicholl} et~al.,}{{Nicholl}
  et~al.}{2015}]{nicholl15}
{Nicholl} M.,  et~al., 2015, \mn@doi [\mnras] {10.1093/mnras/stv1522}, \href
  {http://adsabs.harvard.edu/abs/2015MNRAS.452.3869N} {452, 3869}

\bibitem[\protect\citeauthoryear{{Nicholl} et~al.,}{{Nicholl}
  et~al.}{2016a}]{nicholl16}
{Nicholl} M.,  et~al., 2016a, \mn@doi [\apj] {10.3847/0004-637X/826/1/39},
  \href {http://adsabs.harvard.edu/abs/2016ApJ...826...39N} {826, 39}

\bibitem[\protect\citeauthoryear{{Nicholl} et~al.,}{{Nicholl}
  et~al.}{2016b}]{nicholl16c}
{Nicholl} M.,  et~al., 2016b, \mn@doi [\apjl] {10.3847/2041-8205/828/2/L18},
  \href {http://adsabs.harvard.edu/abs/2016ApJ...828L..18N} {828, L18}

\bibitem[\protect\citeauthoryear{{Nicholl}, {Berger}, {Margutti}, {Blanchard},
  {Milisavljevic}, {Challis}, {Metzger}  \& {Chornock}}{{Nicholl}
  et~al.}{2017}]{nicholl16b}
{Nicholl} M.,  {Berger} E.,  {Margutti} R.,  {Blanchard} P.~K.,
  {Milisavljevic} D.,  {Challis} P.,  {Metzger} B.~D.,   {Chornock} R.,  2017,
  \mn@doi [\apjl] {10.3847/2041-8213/aa56c5}, \href
  {http://adsabs.harvard.edu/abs/2017ApJ...835L...8N} {835, L8}

\bibitem[\protect\citeauthoryear{{Pastorello} et~al.,}{{Pastorello}
  et~al.}{2010}]{pastorello10}
{Pastorello} A.,  et~al., 2010, \mn@doi [\apjl] {10.1088/2041-8205/724/1/L16},
  \href {http://adsabs.harvard.edu/abs/2010ApJ...724L..16P} {724, L16}

\bibitem[\protect\citeauthoryear{{Perley} et~al.,}{{Perley}
  et~al.}{2016}]{perley16}
{Perley} D.~A.,  et~al., 2016, \mn@doi [\apj] {10.3847/0004-637X/830/1/13},
  \href {http://adsabs.harvard.edu/abs/2016ApJ...830...13P} {830, 13}

\bibitem[\protect\citeauthoryear{{Quimby}, {Aldering}, {Wheeler},
  {H{\"o}flich}, {Akerlof}  \& {Rykoff}}{{Quimby} et~al.}{2007}]{quimby07}
{Quimby} R.~M.,  {Aldering} G.,  {Wheeler} J.~C.,  {H{\"o}flich} P.,  {Akerlof}
  C.~W.,   {Rykoff} E.~S.,  2007, \mn@doi [\apjl] {10.1086/522862}, \href
  {http://adsabs.harvard.edu/abs/2007ApJ...668L..99Q} {668, L99}

\bibitem[\protect\citeauthoryear{{Quimby} et~al.,}{{Quimby}
  et~al.}{2011}]{quimby11}
{Quimby} R.~M.,  et~al., 2011, \mn@doi [\nat] {10.1038/nature10095}, \href
  {http://adsabs.harvard.edu/abs/2011Natur.474..487Q} {474, 487}

\bibitem[\protect\citeauthoryear{{Quimby}, {Yuan}, {Akerlof}  \&
  {Wheeler}}{{Quimby} et~al.}{2013}]{quimby13}
{Quimby} R.~M.,  {Yuan} F.,  {Akerlof} C.,   {Wheeler} J.~C.,  2013, \mn@doi
  [\mnras] {10.1093/mnras/stt213}, \href
  {http://adsabs.harvard.edu/abs/2013MNRAS.431..912Q} {431, 912}

\bibitem[\protect\citeauthoryear{{R{\'e}my-Ruyer} et~al.,}{{R{\'e}my-Ruyer}
  et~al.}{2014}]{remy14}
{R{\'e}my-Ruyer} A.,  et~al., 2014, \mn@doi [\aap]
  {10.1051/0004-6361/201322803}, \href
  {http://adsabs.harvard.edu/abs/2014A%26A...563A..31R} {563, A31}

\bibitem[\protect\citeauthoryear{{SDSS Collaboration} et~al.,}{{SDSS
  Collaboration} et~al.}{2016}]{sdss13}
{SDSS Collaboration} et~al., 2016, preprint, \href
  {http://adsabs.harvard.edu/abs/2016arXiv160802013S} {} (\mn@eprint {arXiv}
  {1608.02013})

\bibitem[\protect\citeauthoryear{{Schlafly} \& {Finkbeiner}}{{Schlafly} \&
  {Finkbeiner}}{2011}]{dustmap}
{Schlafly} E.~F.,  {Finkbeiner} D.~P.,  2011, \mn@doi [\apj]
  {10.1088/0004-637X/737/2/103}, \href
  {http://adsabs.harvard.edu/abs/2011ApJ...737..103S} {737, 103}

\bibitem[\protect\citeauthoryear{{Scovacricchi}, {Nichol}, {Bacon}, {Sullivan}
  \& {Prajs}}{{Scovacricchi} et~al.}{2016}]{scova16}
{Scovacricchi} D.,  {Nichol} R.~C.,  {Bacon} D.,  {Sullivan} M.,   {Prajs} S.,
  2016, \mn@doi [\mnras] {10.1093/mnras/stv2752}, \href
  {http://adsabs.harvard.edu/abs/2016MNRAS.456.1700S} {456, 1700}

\bibitem[\protect\citeauthoryear{{Skrutskie} et~al.,}{{Skrutskie}
  et~al.}{2006}]{skrutskie06}
{Skrutskie} M.~F.,  et~al., 2006, \mn@doi [\aj] {10.1086/498708}, \href
  {http://adsabs.harvard.edu/abs/2006AJ....131.1163S} {131, 1163}

\bibitem[\protect\citeauthoryear{{Taddia} et~al.,}{{Taddia}
  et~al.}{2016}]{taddia16}
{Taddia} F.,  et~al., 2016, \mn@doi [\aap] {10.1051/0004-6361/201628703}, \href
  {http://adsabs.harvard.edu/abs/2016A%26A...592A..89T} {592, A89}

\bibitem[\protect\citeauthoryear{{Valenti} et~al.,}{{Valenti}
  et~al.}{2011}]{quba}
{Valenti} S.,  et~al., 2011, \mn@doi [\mnras]
  {10.1111/j.1365-2966.2011.19262.x}, \href
  {http://adsabs.harvard.edu/abs/2011MNRAS.416.3138V} {416, 3138}

\bibitem[\protect\citeauthoryear{{Valenti} et~al.,}{{Valenti}
  et~al.}{2012}]{valenti12}
{Valenti} S.,  et~al., 2012, \mn@doi [\apjl] {10.1088/2041-8205/749/2/L28},
  \href {http://adsabs.harvard.edu/abs/2012ApJ...749L..28V} {749, L28}

\bibitem[\protect\citeauthoryear{{Woosley}}{{Woosley}}{2010}]{woosley10}
{Woosley} S.~E.,  2010, \mn@doi [\apjl] {10.1088/2041-8205/719/2/L204}, \href
  {http://adsabs.harvard.edu/abs/2010ApJ...719L.204W} {719, L204}

\bibitem[\protect\citeauthoryear{{Yan} et~al.,}{{Yan} et~al.}{2016}]{yan16}
{Yan} L.,  et~al., 2016, preprint, \href
  {http://adsabs.harvard.edu/abs/2016arXiv161102782Y} {} (\mn@eprint {arXiv}
  {1611.02782})

\bibitem[\protect\citeauthoryear{{van Dokkum}}{{van Dokkum}}{2001}]{lacosmic}
{van Dokkum} P.~G.,  2001, \mn@doi [\pasp] {10.1086/323894}, \href
  {http://adsabs.harvard.edu/abs/2001PASP..113.1420V} {113, 1420}

\makeatother
\end{thebibliography}

% Don't change these lines
\bsp	% typesetting comment
\label{lastpage}
\end{document}